\documentclass[english,journal]{IEEEtran}
\usepackage[T1]{fontenc}
\usepackage[latin9]{luainputenc}
\usepackage{array}
\usepackage{float}
\usepackage{booktabs}
\usepackage{mathtools}
\usepackage{multirow}
\usepackage{amsmath}
\usepackage{amssymb}
\usepackage{graphicx}

\makeatletter

\providecommand{\tabularnewline}{\\}

\usepackage{arydshln}
\date{}
\usepackage{etoolbox}
\usepackage[acronym,toc,shortcuts]{glossaries}
\usepackage{amsmath,amsfonts,amsthm,bm} 
\usepackage{mathtools,amssymb,lipsum}

\usepackage{cuted}
    \newlength{\halfpagewidth}
    \divide\halfpagewidth by 1
    
\allowdisplaybreaks
\usepackage[table]{xcolor}

\usepackage{soul}
\soulregister\cite7 
\soulregister\citep7 
\soulregister\citet7 
\soulregister\ref7 
\soulregister\pageref7 
\usepackage{color, xcolor}
\usepackage{diagbox}

\makeatother

\usepackage{babel}
\begin{document}
\title{Physics-Informed Neural Networks for Prognostics and Health Management
of Lithium-Ion Batteries}
\author{Pengfei~Wen, Zhi-Sheng~Ye,~\IEEEmembership{Senior Member,~IEEE}, Yong~Li,
Shaowei~Chen,~\IEEEmembership{Member,~IEEE},    Pu~Xie, and    Shuai~Zhao,~\IEEEmembership{Member,~IEEE}
\thanks{Pengfei Wen is with the School of Electronics and Information, Northwestern Polytechnical University, Xi'an 710072, China. He is also with the Department of Industrial Systems Engineering and Management, National University of Singapore, Singapore 119077, Singapore. (Email: wenpengfei@mail.nwpu.edu.cn; pengfei\_wen@u.nus.edu)}
\thanks{Zhi-Sheng Ye is with the Department of Industrial Systems Engineering and Management, National University of Singapore, Singapore 119077, Singapore. (Email: yez@nus.edu.sg)}
\thanks{Yong Li and Shaowei Chen are with the School of Electronics and Information, Northwestern Polytechnical University, Xi'an 710072, China. (Email: ruikel@nwpu.edu.cn; cgong@nwpu.edu.cn)}
\thanks{Pu Xie is with the Department of Aeronautics and Astronautics, Stanford University, Stanford, California, USA, 94305. (Email: xiepu@stanford.edu)}
\thanks{Shuai Zhao is with the AAU Energy, Aalborg University, Aalborg 9220, Denmark. (Email: szh@energy.aau.dk)
(\textit{Corresponding author: Shuai Zhao})}}
\maketitle
\begin{abstract}
For Prognostics and Health Management (PHM) of Lithium-ion (Li-ion)
batteries, many models have been established to characterize their
degradation process. The existing empirical or physical models can
reveal important information regarding the degradation dynamics. However,
there are no general and flexible methods to fuse the information
represented by those models. Physics-Informed Neural Network (PINN)
is an efficient tool to fuse empirical or physical dynamic models
with data-driven models. To take full advantage of various information
sources, we propose a model fusion scheme based on PINN. It is implemented
by developing a semi-empirical semi-physical Partial Differential
Equation (PDE) to model the degradation dynamics of Li-ion batteries.
When there is little prior knowledge about the dynamics, we leverage
the data-driven Deep Hidden Physics Model (DeepHPM) to discover the
underlying governing dynamic models. The uncovered dynamics information
is then fused with that mined by the surrogate neural network in the
PINN framework. Moreover, an uncertainty-based adaptive weighting
method is employed to balance the multiple learning tasks when training
the PINN. The proposed methods are verified on a public dataset of
Li-ion Phosphate (LFP)/graphite batteries.
\end{abstract}


\section{Introduction}

\IEEEPARstart{G}{lobal} sales of Electric Vehicles (EVs) doubled
in 2021 from 2020 to a new record of 6.6 million according to \cite{RN405}
released by International Energy Agency (IEA). It keeps rising strongly
and 2 million were sold in the first quarter of 2022, up 75\% from
the same period in the previous year. As their power source, Lithium-ion
(Li-ion) batteries have become one of the most important industrial
consumables. The demand for batteries reached 340 Gigawatt-hours (GWh)
in 2021, which also doubled from the previous year. In the same year,
the average battery price is USD 132/kWh. The cost of battery replacement
due to its degradation is still high \cite{RN475}, and the Battery
Management System (BMS) is developed to perform Prognostics and Health
Management (PHM). The primary tasks of BMS include both State of Health
(SoH) estimation and Remaining Useful Life (RUL) prognostics \cite{RN477}.
Fused usage information collected from various sources via BMS is
expected to significantly improve the performance of PHM.

Generally, information fusion has been achieved in various flexible
forms, including data-level , feature-level, decision-level, and model-level.
The prognostic models can be categorized as physics-based models,
experience-based models, and data-driven models \cite{RN141,RN408}.
Physics-based and experience-based models represent abundant prior
domain knowledge of the monitored systems condensed by experts that
have been widely accepted \cite{RN476}. There is no insurmountable
gap between these two categories of models since many physics models
originate from empirical models, such as the Paris--Erdogan equation.
On the contrary, building data-driven models is a process of mining
latent information in data. Dynamic models are commonly built to represent
the governing dynamics during degrading, which can provide critical
insights into the internal change of monitored systems.

According to the No-Free-Lunch (NFL) theorem, information from different
models fits distinct problems well. Model fusion can leverage the
advantages of combining the information from different categories
of available models for PHM. It can be difficult to categorize fusion
forms of model fusion. Five mostly-applied forms of fusing models
are reviewed according to the combination and interfaces of distinct
types of prognostic models \cite{RN141}. Fusing physics-based models
and data-driven PHM methods drags much attention \cite{RN204}. Among
these fusion forms, transition equations built based on physics or
experience have been widely integrated into the framework of Bayesian
filtering \cite{RN205,RN130}. These transition equations are then
used to iterate the degradation state. This implementation of fusing
physics-informed dynamic models and data-driven prognostic models
inspires the frontiers of Physics-Informed Machine Learning (PIML)
\cite{RN362} in PHM.

The principle of PIML is to fuse the physics (experience)-based models
and data-driven models. As reviewed in \cite{RN141}, this category
of model fusion can be instantiated in various forms as well. Typical
implementations of PIML are reviewed in \cite{RN362} in terms of
their applicable scenarios, principles, frameworks, and applications.
As a formalized framework among PIML methods, Physics-Informed Neural
Network (PINN) gains momentum since dynamic models established in
various forms can be integrated. This high-level flexibility makes
it popular in multiple frontiers such as epidemiology \cite{RN287},
power electronics \cite{RN400,RN401}, acoustics \cite{RN406}, and
electromagnetics \cite{RN407}. For instance, \cite{RN287} showed
how PINN is capable of forecasting various highly-infectious diseases
progression (e.g. COVID, HIV, and Ebola, etc.) whose spread can be
modeled by Ordinary Differential Equations (ODEs) or Partial Differential
Equations  (PDEs). \cite{RN400} monitored the parameter variations
of a DC-DC Buck converter robustly and accurately based on a small
training dataset of only 360 data samples with PINN. These concrete
examples in different fields show the significance and generalizability
of this method.

In this paper, we propose a semi-physics and semi-empirical dynamic
model based on the Verhulst model \cite{RN278} to capture the fading
trends of the battery SoH. This model is further generalized as a
PDE considering observable features given charging and discharging
profiles, and a new parameter that represents the level of initial
Solid Electrolyte Interphase (SEI) formation is introduced. To estimate
the SoH of Li-ion batteries, we introduce PINN to fuse the prior information
formulated as the dynamic model and the information extracted from
monitoring data. Besides, a Deep Hidden Physics Model (DeepHPM) is
also employed to discover the dynamics as a comparison with the improved
Verhulst model for SoH estimation. It is also used for RUL prognostics
without any explicit dynamic model. This implementation is a fusion
of information mined by two data-driven models with different functions
\cite{RN141}. During the training of PINN, we use an uncertainty-based
method to adaptively weigh the losses of subsequently proposed multi-task
learning. The proposed methods are then verified on a public experimental
cycling dataset of Li-ion Phosphate (LFP)/graphite batteries.

The contributions of this paper are threefold: 1) We propose a new
scheme to seamlessly fuse prior information of physical or empirical
degradation models and that of monitoring data in PHM for Li-ion batteries.
2) We propose a new semi-physical semi-empirical dynamic degradation
to provide more insight into capacity loss. 3) We propose a scheme
of fusing information of two distinct-designed data-driven models
when there is little prior information. Moreover, the proposed models
can be trained with an adaptive weighting method, which solves the
challenge of combining losses appropriately in training PINNs. The
code and data accompanying this paper are available on GitHub\footnote{{[}Online{]}. Available: https://github.com/WenPengfei0823/PINN-Battery-Prognostics}.

The rest of this article is organized as follows. Section \ref{sec:Related-Work}
provides an overview of related work in terms of degradation modeling
and the schemes of fusing dynamics in Deep Learning (DL). Section
\ref{sec: Dynamic Models} setups the focused problem and the establishment
of two categories of dynamic models is introduced, including an improved
Verhulst model and DeepHPM-based automatically discovered models.
Section \ref{sec: Methodology} presents the framework of PINN for
model fusion and its training method, where an adaptively uncertainty-based
multi-task balancing technique is included. Section \ref{sec: Dataset Description and Processing}
and \ref{sec: Case-Study} verify the proposed methods on a public
experimental dataset, and several details are provided in Appendix
\ref{sec: Hyper-Parameters-Tuning}. Finally, Section \ref{sec:Conclusion}
concludes this article.

\section{Related Work\label{sec:Related-Work}}

\subsection{Dynamic Degradation Models of Li-Ion-Batteries}

Dynamic models for Li-ion-batteries degradation can be established
based on the classic Pseudo-Two-dimensional (P2D) model, which possesses
high accuracy but is severely restricted due to the subsequent complexity
and computation burden \cite{RN403}. The Single-Particle (SP) model
is then proposed to simplify the P2D model, which assumes that both
electrodes consist of multiple uniform-sized spherical particles \cite{RN282}.
Based on the SP model, a widely-accepted degradation mechanism was
proposed. Since electrolytes can be reduced in the presence of lithiated
carbon, SEI formation is caused by the lithium carbonate. The newly
exposed surface of the particles is then covered by SEI during the
cycling. Under this mechanism, the degradation rate considering both
Diffusion Induced Stresses (DIS), crack growth, and SEI thickness
growth was derived in \cite{RN285,RN282}. There are many parameters
to estimate and verify in this dynamic model proposed fully based
on prior electrochemical knowledge, which restricts its generalizability.

Another category of degradation models focuses more on the utilization
of available degradation data with monitoring time or cycles, and
then the degradation trends are modeled by analytical functions under
the given statistical assumptions \cite{RN404}. These models commonly
oversimplify the physics that the degradation trends of Li-ion batteries
should follow. Semi-physics and semi-empirical dynamic models are
proposed to take both factors into account \cite{RN277}. For instance,
an exponential-law model can be derived based on battery Coulombic
efficiency \cite{RN272}. This exponential model can also be integrated
from an empirically given differential form and then its parameters
are estimated based on temperature stress, State of Charge (SoC) stress,
time stress, and Depth of Discharge (DoD) stress \cite{RN277}.

In view of the fact that there is an upper bound for the capacity
loss, the Verhulst model has been proposed to quantify the rate of
capacity loss \cite{RN278}. As can be seen, there usually exists
more or less available information from physics for SoH estimation,
while this is not always true for RUL prognostics. To fill this gap,
discovery for dynamic models with a preset fixed form or completely
arbitrary form has been explored by developing PDE-Net \cite{RN398,RN399}
and DeepHPM \cite{RN350,RN366} respectively. These models discover
PDEs by training specialized layers. In PDE-Net, convolution kernels
are trained to approximate the differential operators, and other network
parameters are trained to approximate finite-order linear PDEs. Discrete-time
models can be accordingly constructed. In DeepHPM, the most significant
difference is that it uses the Automatic Differentiator (AutoDiff)
to compute the analytical differentiation. As a result, DeepHPM can
handle both discrete and continuous, linear and nonlinear models.

\subsection{Fusion of the Dynamics in Deep Learning}

In the literature, several studies have been conducted integrating
the dynamic models into DL, as follows.

Difference equations are the most commonly used discrete-time models.
For example, temporal-modeling-specialized Recurrent Neural Network
(RNN) and its variants can be applied to quantify the dynamics between
different monitoring times by using a series of equations \cite{RN479}.
Several classic variants of RNN have fixed forms of these equations,
such as the Long-Short Term Memory (LSTM) NN and the Gated Recurrent
Unit (GRU) NN. Based on the domain knowledge of monitored systems,
these equations can be re-designed to fuse prior physical information.
Considering the cumulative damage of the grease and bearing damage,
the physics of wind turbine main bearing fatigue was used to design
a specific RNN cell for its PHM \cite{RN359}. More physics including
grease curves, bearing design data, and bearing design curves were
further fused in their following work \cite{RN370}. Since the RNN
cell was completely reconstructed, these methods demand quite a deep
understanding of the monitoring systems. Otherwise, these physics-informed
cells may be inferior to the generic LSTM or GRU architectures.

Compared with these methods, the constraint learning scheme is more
convenient to incorporate physical dynamics. It learns the knowledge
by adding regularization that is determined by the physical constraints.
For the inevitable calculation of differential terms in the dynamic
models, approximating by the specially designed filters and AutoDiff
are two general methods. For example, to approximate the spatial gradients
in computer vision, the Sobel filter is an efficient approximator
\cite{RN397}. By contrast, PINN uses AutoDiff that can provide the
exact partial derivatives. This scheme can handle both discrete and
continuous time models.

When building the discrete-time models by using PINN, a multi-outputs
NN is established where each neuron on the output layer outputs the
predicted implicit intermediate states between two observable ones.
These outputs are coupled and constrained in the framework of the
implicit Runge-Kutta method during training \cite{RN289,RN400}. In
this way, a priori-designed Butcher tableau is required given a certain
number of the setup intermediate states coupled with a specific interval
length between the two observable states \cite{RN402}. On the contrary,
an NN is established to directly approximate the implicit solution
of the continuous time dynamic models.

In this paper, we focus on the continuous form of PINN to fuse the
established dynamic models, which are provided by using the Verhulst
model and the DeepHPM corresponding to the cases whether the prior
domain knowledge is partially known or not at all.

\section{Problem Statement\label{sec: Dynamic Models}}

Battery aging results from a combined action of both times went and
cycling \cite{RN277,RN374}. Fading in a Li-ion battery upon charging
and discharging can be attributed to a coupled mechanical-chemical
degradation within the cell. The DISs can boost the growth of the
cracks on the electrode surfaces during cycling and thus leading to
the growth of SEI \cite{RN282,RN285}. This degradation mechanism
assembles the fatigue process of materials upon cyclic loading \cite{RN277},
where the capacity loss due to the stress accumulates independently
and ultimately causes life loss. Such a process can be categorized
as cycle aging. At the same time, a battery also degrades over time
because of the inherent slow electrochemical reaction, forming the
calendar aging. Since charging and discharging processes can often
be completed in dozens of minutes while calendar aging becomes observable
usually on the scales of months or even years, cycle aging contributes
the most to capacity and life loss. As a result, we assume that calendar
aging can be neglected here when modeling the degradation dynamics
of batteries.

The failure time of a Li-ion battery is commonly defined as the time
when its currently usable capacity reduces below a pre-specified threshold
\cite{RN474}. SoH of a battery is consequently related to its current
capacity and is typically defined as the ratio of current capacity
over the nominal one \cite{RN269}. Here we focus on the Percentage
Capacity Loss (PCL) $u$ which represents the relative gap between
the usable capacity and the nominal one, and it is denoted as:
\begin{equation}
u_{k}=1-SoH_{k}=1-\frac{Q_{k}}{Q_{Nom}}\times100\%,\label{eq: PCL and SoH}
\end{equation}
where $Q_{Nom}$ represents the nominal capacity and $Q_{k}$ represents
the usable capacity in $k^{\mathrm{th}}$ cycle.

\subsection{Dynamic Models}

Without loss of generality, PCL $u$ can be set as a uni-variate function
of a virtual continuous independent time variable $t$ \cite{RN57}
whose unit is still the cycles. When a battery is operated upon identical
conditions upon each cycle, a series of observations can be acquired
at discrete time $t_{k}$:
\begin{equation}
u_{k}=u\left(t_{k}\right).
\end{equation}
To handle the degradation dynamics of the battery degradation trend,
the fading rate of its capacity can be denoted as
\begin{equation}
\frac{\mathrm{d}u\left(t\right)}{\mathrm{d}t}=\mathcal{G}\left(t,u;\Theta\right).\label{eq: Basic ODE}
\end{equation}
Eq. (\ref{eq: Basic ODE}) is an explicit ODE parameterized by set
$\Theta$, and $\mathcal{G}$ denotes a nonlinear function of $t$
and $u$. For instance, based on SEI formation and the growth process
on both the initial and cracked surface \cite{RN285,RN282}, eq. (\ref{eq: Basic ODE})
can be specified as
\begin{align}
\left.\frac{\mathrm{d}u\left(t\right)}{\mathrm{d}t}\right|_{t=t_{k}} & =\theta_{1}\theta_{5}\left(1+\theta_{2}t_{k}\right)^{\frac{\theta_{3}}{2-\theta_{3}}}+\theta_{4}t_{k}^{-\frac{1}{2}},\label{eq: Single particle ODE}\\
 & +\theta_{1}\theta_{6}\sum_{l=1}^{k-1}\left[1+\theta_{2}\left(t_{k}-t_{l}\right)\right]^{\frac{\theta_{3}}{2-\theta_{3}}}\left(t_{k}-t_{l}\right)^{-\frac{1}{2}},\nonumber \\
l & <k,
\end{align}
where $t_{l}$ represents the monitoring time earlier than $t_{k}$
and  $\Theta=\left\{ \theta_{1},\theta_{2},\ldots,\theta_{6}\right\} $
are composite parameters. Each of them should be set according to
dozens of electrochemical parameters including the geometric area
of the graphite electrode film, the activation energy for crack propagation,
and the solid phase porosity of the electrode film, etc. This theoretical
model details molecular-level aging mechanism but it can be hardly
employed in practical applications due to the lack of detailed cell
conditions. Such a dynamic model describes a degradation trend reported
in \cite{RN388} with a decelerated trend of $u$ during the early
cycles and a moderate linear trend \cite{RN252} during the latter
cycles. Based on the observed trend, a simplified semi-empirical model
was proposed in \cite{RN277} by using regression analysis:
\begin{equation}
\frac{\mathrm{d}u\left(t\right)}{\mathrm{d}t}=\theta\left[1-u\left(t\right)\right],\label{eq: Xu's model}
\end{equation}
where $\theta$ denotes a basic linearized degradation rate that is
co-determined by SoC, DoD, and cell temperature. An exponential-function
solution with a decreasing absolute value of derivative can be acquired
by integrating (\ref{eq: Xu's model}) to $u$. It can be observed
that this model may not fit the battery degradation with an accelerated
trend during the early cycles reported in \cite{RN354,RN385}.

\subsection{Improved Verhulst Dynamic Model}

Here the function $\mathcal{G}\left(t,u;\Theta\right)$ is also made
dependent on the health state of the cell, and takes the simplest
linear form as an instance:
\begin{align}
\frac{\mathrm{d}u\left(t\right)}{\mathrm{d}t} & =ru\left(t\right),\label{eq: Exponential model}\\
\mathrm{\mathit{s.t.}}\,\,u\left(t\right),\,r & >0.\nonumber 
\end{align}
where $r$ is the degradation constant. Eq. (\ref{eq: Exponential model})
and partial term of (\ref{eq: Xu's model}) mark that the rate at
which the capacity loss changes is proportional to the usable capacity
at that time. The solution of (\ref{eq: Exponential model}) is the
exponential function $u\left(t\right)=u_{0}\exp\left(rt\right)$ with
an initial loss $u_{0}$, representing an accelerated degradation
trend. This initial loss $u_{0}$ is commonly set as 10\% \cite{RN285}.
Considering the capacity loss results from SEI growth and so on is
always finite, a constant $K$ representing the upper bound of capacity
loss is introduced to constrain the increasing rate of loss:
\begin{align}
\frac{\mathrm{d}u\left(t\right)}{\mathrm{d}t} & =ru\left(t\right)\left[1-\frac{u\left(t\right)}{K}\right],\label{eq: Verhulst model}\\
\mathrm{s.t.}\,\,r & >0,\nonumber \\
0 & <u\left(t\right)<K.\nonumber 
\end{align}
Eq. (\ref{eq: Verhulst model}) is also known as the logistic differential
equation proposed by Pierre Verhulst to model the population growth
process \cite{RN278}. When the lost capacity increases close to $K$,
the increasing rate $\mathrm{d}u\left(t\right)/\mathrm{d}t$ will
decline until it is close to 0. Since a battery is commonly considered
as failed when its $u$ increases exceed 20\%, it can be \textit{a
priori} roughly known that $K$ has a range of 20\% to 100\%. Furthermore,
SEI will naturally form after a brand-new battery is manufactured
and before use. In this process, the capacity loss caused by the SEI
formation is supposed not to be governed by these dynamic models.
Instead, a parameter $C$ that denotes such initial capacity loss
is introduced to modify the model (\ref{eq: Verhulst model}) as
\begin{align}
\frac{\mathrm{d}u\left(t\right)}{\mathrm{d}t} & =r\left[u\left(t\right)-C\right]\left[1-\frac{u\left(t\right)-C}{K-C}\right].\label{eq: Modified Verhulst model}\\
\mathrm{s.t.}\,\,r & >0,\nonumber \\
0 & <u\left(t\right)<K,\nonumber \\
0 & <C\leq u_{0}.\nonumber 
\end{align}
Even though under identical operation conditions, different cells
manufactured in the same batch may show distinct degradation characteristics.
With the setup that $u$ is a uni-variate function of $t$, this cell-to-cell
heterogeneity can be modeled by the difference in degradation model
parameters \cite{RN389,RN66}. Considering the sole time-independent
variable $t$ cannot distinguish specific trajectory when different
batteries are fading, other variables that can indicate the latent
health state can be introduced. In this setup, $u$ is expanded to
a multi-variate function of both $\boldsymbol{x}=\left[x_{1},x_{2},\ldots,x_{S}\right]^{\mathrm{T}}\in\mathbb{R}^{S}$
and $t$. An $S$-dimensional Health Indicator (HI) $\boldsymbol{x}$
can characterize the health state during the battery fading, and both
monitoring data and designed representative features can act as health
indicators \cite{RN252,RN390}. It is feasible that cell-to-cell heterogeneity
is quantified by different specific combinations of values in the
feature space. The degradation rate given $\boldsymbol{x}$ and monitoring
time $t$ is subsequently modeled by a PDE as:
\begin{align}
\frac{\mathrm{\partial}u\left(\boldsymbol{x},t\right)}{\mathrm{\partial}t} & =r\left[u\left(\boldsymbol{x},t\right)-C\right]\left[1-\frac{u\left(\boldsymbol{x},t\right)-C}{K-C}\right].\label{eq: PDE Verhulst model}\\
\mathrm{s.t.}\,\,r & >0,\nonumber \\
0 & <u\left(t\right)<K,\nonumber \\
0 & <C\leq u_{0}.\nonumber 
\end{align}
The parameters of eq. (\ref{eq: PDE Verhulst model}) are of the same
meanings as those in eq. (\ref{eq: Modified Verhulst model}).

\subsection{Data-Driven Dynamic Model\label{subsec: DeepHPM}}

The dynamic model (\ref{eq: PDE Verhulst model}) depicts one of many
forms of SoH degrading. Beyond SoH prognostics, it can be difficult
to distill degradation mechanisms when predicting other variables
such as RUL. Following (\ref{eq: Basic ODE}), we define more generalized
nonlinear dynamics parameterized by $\Theta$ to distill the mechanisms
governing the evolution \cite{RN350} of given data of HIs and time
as
\begin{equation}
u_{t}-\mathcal{G}\left(\boldsymbol{x},t,u,u_{\boldsymbol{x}},u_{\boldsymbol{x}\boldsymbol{x}},u_{\boldsymbol{x}\boldsymbol{x}\boldsymbol{x}},\ldots;\Theta\right)=0.\label{eq: Basic PDE}
\end{equation}
In eq. (\ref{eq: Basic PDE}), $u_{\boldsymbol{x}}=\left[\frac{\mathrm{\partial}u}{\mathrm{\partial}x_{1}},\frac{\mathrm{\partial}u}{\mathrm{\partial}x_{2}},\ldots,\frac{\mathrm{\partial}u}{\mathrm{\partial}x_{S}}\right]^{\mathrm{T}}$
denotes the first-order partial derivative of $u$ with respect to
$\boldsymbol{x}$ and so on. The nonlinear function $\mathcal{G}$
poses more flexible relations on $t$, $u$, and their any-order partial
derivatives. Subsequently, an infinite dimensional dynamical system
can be represented by $\mathcal{G}$ \cite{RN350}. As can be seen,
(\ref{eq: PDE Verhulst model}) is a specific implementation of (\ref{eq: Basic PDE})
where $\mathcal{G}\left(u;\Theta\right)=r\left(u-C\right)\left(1-\frac{u-C}{K-C}\right)$.
In most cases of health monitoring, it is challenging to define an
explicit dynamic model. With scattered and noisy monitoring observations,
bias still exists in the PDE model (\ref{eq: PDE Verhulst model})
and the more-complicated latent dynamics. We construct a Neural Network
(NN) as a function approximator \cite{RN391} to fill this gap beyond
a particular family of basis functions \cite{RN366}, forming a DeepHPM
\cite{RN350}:
\begin{equation}
u_{t}-\mathrm{DeepHPM}\left(\boldsymbol{x},t,u,u_{\boldsymbol{x}},u_{\boldsymbol{x}\boldsymbol{x}},u_{\boldsymbol{x}\boldsymbol{x}\boldsymbol{x}},\ldots;\Theta\right)=0.\label{eq: DeepHPM model}
\end{equation}
Here we denote the NN used to approximate $\mathcal{G}$ as the DeepHPM.
The parameter set $\Theta$ represents the trainable network parameters
of the function approximator DeepHPM.

\section{Methodology\label{sec: Methodology}}

The PDE dynamic model established based on prior knowledge (\ref{eq: PDE Verhulst model})
or approximated by NN (\ref{eq: DeepHPM model}) can be quite hard
to solve. Based on the well-known capability of NNs as universal function
approximators, we employ another NN parameterized by $\Phi$ to approximate
the hidden solution $u\left(\boldsymbol{x},t;\Phi\right)$ of the
system. Solution $u\left(\boldsymbol{x},t;\Phi\right)$ is also called
the \textit{surrogate} network. With this NN solver, the direct access
or approximations to the involved partial derivatives are unnecessary
\cite{RN289,RN350}. Model fusion of the surrogate NN with the explicit
PDE models or DeepHPM formulates a PINN.

\subsection{Model Fusion by PINN}

A typical structure of PINN consists of three modules including a
dynamic model, a surrogate NN, and an automatic differentiator. The
dynamic model $\mathcal{G}\left(\boldsymbol{x},t,u,u_{\boldsymbol{x}},u_{\boldsymbol{x}\boldsymbol{x}},u_{\boldsymbol{x}\boldsymbol{x}\boldsymbol{x}},\ldots;\Theta\right)$
distills the mechanisms governing the dynamics of a degrading system,
which can be either explicitly defined as (\ref{eq: PDE Verhulst model})
or approximated by a DeepHPM as (\ref{eq: DeepHPM model}). The surrogate
NN $u\left(\boldsymbol{x},t;\Phi\right)$ is used to approximate the
hidden solution $u\left(\boldsymbol{x},t\right)$ of the dynamic model,
and the AutoDiff \cite{RN392} is used to calculate values of all
involved partial differentials input to the dynamic model.

Except for solving the latent $u\left(\boldsymbol{x},t\right)$ to
build the prognostic model, discovering the dynamic model is realized
by identifying the parameter set $\Theta$ of either specific PDE
or DeepHPM. Identifying the unknown parameters forms a high-dimensional
inverse problem describing the dynamical system. With the surrogate
NN $u\left(\boldsymbol{x},t;\Phi\right)$, we define the left-hand
side of eq. (\ref{eq: DeepHPM model}) as a function $f\left(\boldsymbol{x},t;\Phi,\Theta\right)$:
\begin{equation}
f\left(\boldsymbol{x},t;\Phi,\Theta\right)\coloneqq u_{t}\left(\boldsymbol{x},t;\Phi\right)-\mathcal{G}\left(\boldsymbol{x},t,u,u_{\boldsymbol{x}},u_{\boldsymbol{x}\boldsymbol{x}},u_{\boldsymbol{x}\boldsymbol{x}\boldsymbol{x}},\ldots;\Theta\right).
\end{equation}
The parameters $\Phi$ of surrogate NN $u\left(\boldsymbol{x},t;\Phi\right)$
and $\Theta$ of dynamic model $\mathcal{G}\left(\boldsymbol{x},t,u,u_{\boldsymbol{x}},u_{\boldsymbol{x}\boldsymbol{x}},u_{\boldsymbol{x}\boldsymbol{x}\boldsymbol{x}},\ldots;\Theta\right)$
can be trained by minimizing the mean squared error losses:
\begin{equation}
\mathcal{L}=\lambda_{u}\mathcal{L}_{u}+\lambda_{f}\mathcal{L}_{f}+\lambda_{f_{t}}\mathcal{L}_{f_{t}},\label{eq: Loss function}
\end{equation}
where
\begin{align}
\mathcal{L}_{u}= & \sum_{i=1}^{N}\left[u\left(\boldsymbol{x}_{i},t_{i};\Phi\right)-u_{i}\right]^{2},\label{eq: Loss_u}\\
\mathcal{L}_{f}= & \sum_{i=1}^{N}\left[f\left(\boldsymbol{x}_{i},t_{i};\Phi,\Theta\right)\right]^{2},\label{eq: Loss_f}\\
\mathcal{L}_{f_{t}}= & \sum_{i=1}^{N}\left[f_{t}\left(\boldsymbol{x}_{i},t_{i};\Phi,\Theta\right)\right]^{2}.\label{eq: Loss_f_t}
\end{align}
Loss function (\ref{eq: Loss function}) actually represents multi-task
learning. Weight coefficients $\lambda_{u}$, $\lambda_{f}$, and
$\lambda_{f_{t}}$ can be tuned to balance the loss terms in the training
process. The loss term $\mathcal{L}_{u}$ corresponds to the regression
fitting error at the collected observations, while $\mathcal{L}_{f}$
and $\mathcal{L}_{f_{t}}$ enforce the structure of PINN imposed by
(\ref{eq: Basic PDE}) and its first-order partial derivative with
respect to $t$ \cite{RN463}. Here $\mathcal{D}_{Train}=\left\{ \boldsymbol{x}_{i},t_{i},u_{i}\right\} _{i=1}^{N}$
denotes the training data and $u_{i}$ represent the label to be predicted
corresponding to the input data $\left\{ \boldsymbol{x}_{i},t_{i}\right\} $,
which can be actual SoH or RUL in PHM of batteries for the instance.

\subsection{Framework of PINN\label{subsec: Framework of PINN}}

\begin{figure}[h]
\centering{}\includegraphics[width=3.5in]{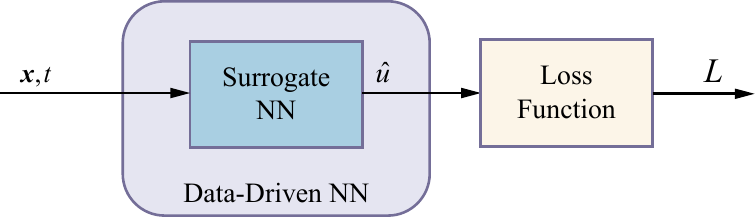}\caption{Framework of typical data-driven vanilla NN.\label{fig: Frameworks-Data driven}}
\end{figure}
\begin{figure*}
\begin{centering}
\includegraphics[width=5in]{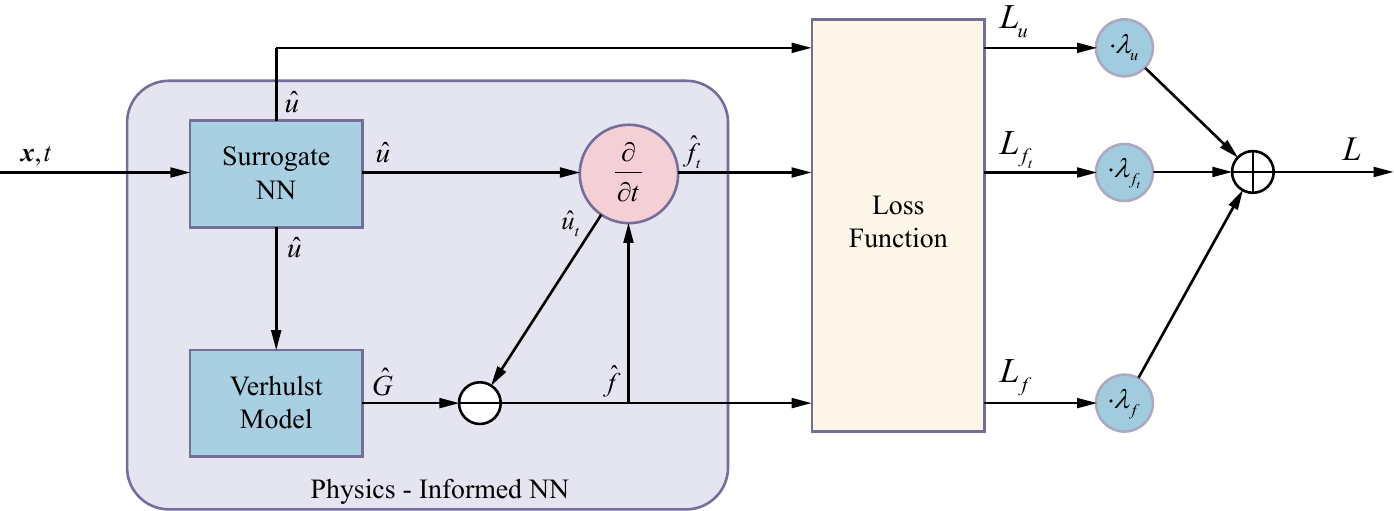}
\par\end{centering}
\caption{Framework of PINN-Verhulst.\label{fig: Frameworks-Verhulst}}
\end{figure*}
\begin{figure*}
\begin{centering}
\includegraphics[width=6.5in]{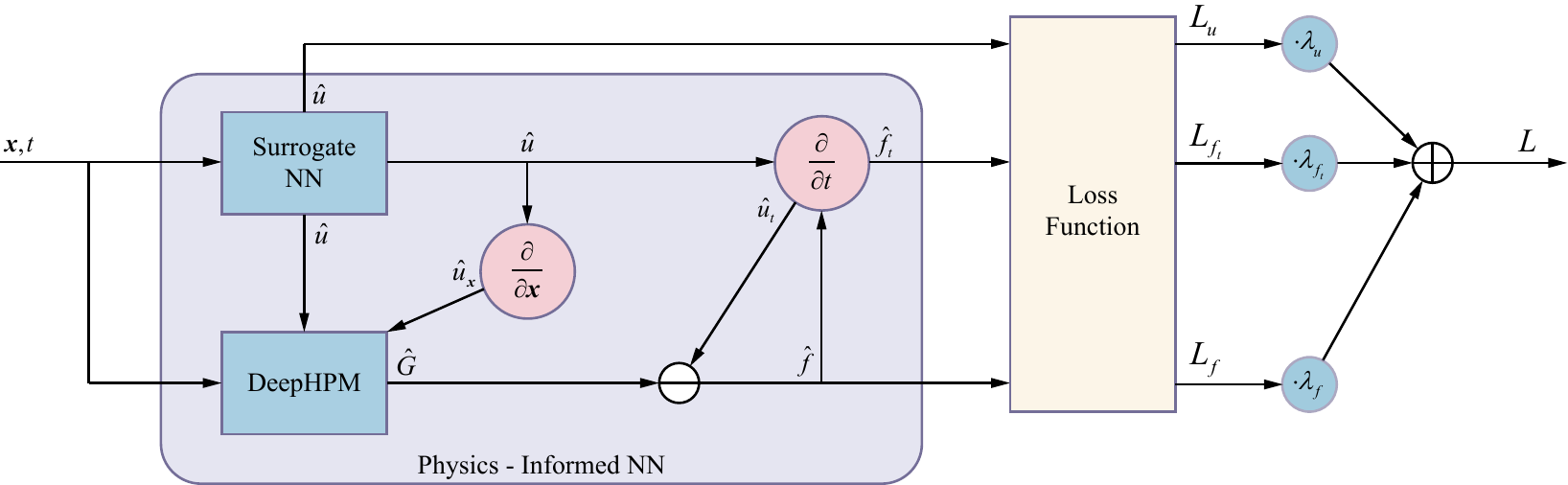}
\par\end{centering}
\caption{Framework of PINN-DeepHPM.\label{fig: Frameworks-DeepHPM}}
\end{figure*}
Here the involved frameworks include a data-driven vanilla NN (baseline),
a PINN with the Verhulst equation as the dynamic model (PINN-Verhulst),
and a PINN with a DeepHPM as the dynamic model (PINN-DeepHPM). These
frameworks are shown in Fig. \ref{fig: Frameworks-Data driven}, \ref{fig: Frameworks-Verhulst},
and \ref{fig: Frameworks-DeepHPM}, respectively. To compare the respective
learning framework clearer, the vanilla NN in the baseline is also
denoted as a surrogate NN although there is no PDE to be solved. Since
DeepHPM is used to discover the dynamic models when there are no available
explicit dynamic models, it can be seen from Fig. \ref{fig: Frameworks-Verhulst}
and \ref{fig: Frameworks-DeepHPM} that those two frameworks are similar.
The key difference is that the Verhulst model in Fig. \ref{fig: Frameworks-Verhulst}
is then substituted as the DeepHPM in Fig. \ref{fig: Frameworks-DeepHPM},
and the corresponding partial derivatives input to the dynamic model
are therefore substituted accordingly. It should be noted that these
inputs are not fixed and need to be further selected in practice.
Moreover, it is demonstrated that the proposed framework is flexible
that can incorporate various forms of physical knowledge.

\begin{figure*}
\begin{centering}
\includegraphics[width=6.5in]{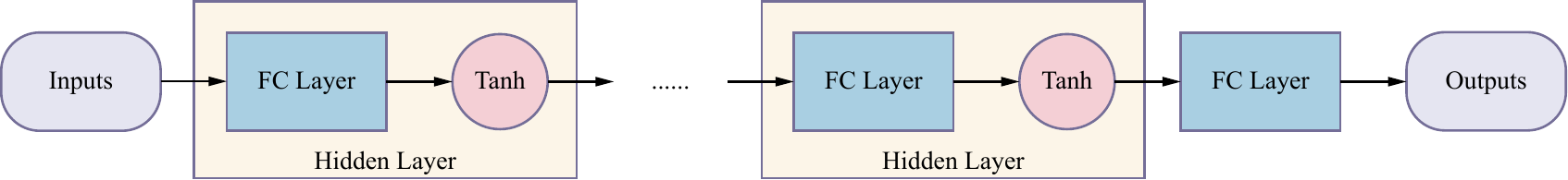}
\par\end{centering}
\caption{Basic structure of NNs.\label{fig: Structure of NNs}}
\end{figure*}

The basic structure of the surrogate NNs is illustrated in Fig. \ref{fig: Structure of NNs}.
The hidden layers of the surrogate NNs are all composed of Fully Connected
(FC) layers, and the hyperbolic tangent is employed as the activation
function due to its differentiability. Structures of all the involved
surrogate NNs in both Fig. \ref{fig: Frameworks-Data driven}, \ref{fig: Frameworks-Verhulst},
and \ref{fig: Frameworks-DeepHPM} are set as this if there are no
special notes. The structure of DeepHPM is set the same as that of
the surrogate NNs.

\subsection{Weight Coefficients Tuning in Training PINN\label{subsec: Weight-Coefficients-Tuning}}

Focusing on the loss function (\ref{eq: Loss function}) of training
a PINN, it can be observed that PINN is constrained or regularized
by the loss imposed by the given set of PDEs beyond a regression task
at the given data set. It is observed that networks are sensitive
to the relative weights of multiple objective functions from desired
training tasks \cite{RN377}. Manually tuning the weighting coefficients
$\lambda$ can be quite expensive, and methods for adaptive tuning
during the training are desirable. Numerical stiffness will cause
unbalanced back-propagated gradients in the PINN training process,
and the weighting coefficients can be accordingly tuned based on the
gradient statistics \cite{RN357}. Other mechanisms have also been
utilized to design the adaptive balancing such as random lookback
\cite{RN376} and homoscedastic aleatoric uncertainty \cite{RN377}.
Following \cite{RN377}, the likelihood of the observed data labels
is assumed as Gaussian. The mean of likelihood is set as the output
of the surrogate NNs and the variance is set according to the weighting
coefficient:
\begin{equation}
p\left(\left.\boldsymbol{u}\right|u\left(\boldsymbol{X},\boldsymbol{t};\Phi\right)\right)=\mathcal{N}\left(u\left(\boldsymbol{X},\boldsymbol{t};\Phi\right),\frac{1}{\lambda_{u}}\right),\label{eq: Likelihood of a task}
\end{equation}
where $\boldsymbol{u}=\left[u_{1},u_{2},\ldots,u_{N}\right]^{\mathrm{T}}$,
$\boldsymbol{X}=\left[\boldsymbol{x}_{1},\boldsymbol{x}_{2},\ldots,\boldsymbol{x}_{N}\right]^{\mathrm{T}}$,
and $\boldsymbol{t}=\left[t_{1},t_{2},\ldots,t_{N}\right]^{\mathrm{T}}$
consist of labels and samples of the training set respectively. The
output corresponding to the loss $\mathcal{L}_{u}$ is used as the
instance and the other objectives in (\ref{eq: Loss function}) can
be accordingly defined. The observation noise scalar is set related
to the weighting coefficient $\lambda$ corresponding to the output
in the loss function. The log-likelihood can be written as 
\begin{equation}
\log p\left(\left.\boldsymbol{u}\right|u\left(\boldsymbol{X},\boldsymbol{t};\Phi\right)\right)\propto-\frac{\lambda_{u}}{2}\left\Vert \boldsymbol{u}-u\left(\boldsymbol{X},\boldsymbol{t};\Phi\right)\right\Vert ^{2}+\frac{1}{2}\log\lambda_{u}.\label{eq: Log-Likelihood of a task}
\end{equation}
Based on (\ref{eq: Loss_u}), (\ref{eq: Loss_f}), (\ref{eq: Loss_f_t}),
and (\ref{eq: Log-Likelihood of a task}), the multi-task minus log-likelihood
can be obtained as (\ref{eq: Log-Likelihood of multiple tasks}).
\begin{figure*}
\begin{align}
 & -\log p\left(\left.\boldsymbol{u},\boldsymbol{f},\boldsymbol{f}_{t}\right|u\left(\boldsymbol{X},\boldsymbol{t};\Phi\right),f\left(\boldsymbol{X},\boldsymbol{t};\Phi,\Theta\right)\right)\nonumber \\
= & -\log\left[p\left(\left.\boldsymbol{u}\right|u\left(\boldsymbol{X},\boldsymbol{t};\Phi\right)\right)\cdot p\left(\left.\boldsymbol{f}\right|f\left(\boldsymbol{X},\boldsymbol{t};\Phi,\Theta\right)\right)\cdot p\left(\left.\boldsymbol{f}_{t}\right|f\left(\boldsymbol{X},\boldsymbol{t};\Phi,\Theta\right)\right)\right],\nonumber \\
= & -\log p\left(\left.\boldsymbol{u}\right|u\left(\boldsymbol{X},\boldsymbol{t};\Phi\right)\right)-\log p\left(\left.\boldsymbol{f}\right|f\left(\boldsymbol{X},\boldsymbol{t};\Phi,\Theta\right)\right)-\log p\left(\left.\boldsymbol{f}_{t}\right|f\left(\boldsymbol{X},\boldsymbol{t};\Phi,\Theta\right)\right),\nonumber \\
= & \lambda_{u}\left\Vert \boldsymbol{u}-u\left(\boldsymbol{X},\boldsymbol{t};\Phi\right)\right\Vert ^{2}+\lambda_{f}\left\Vert \boldsymbol{f}-f\left(\boldsymbol{X},\boldsymbol{t};\Phi,\Theta\right)\right\Vert ^{2}+\lambda_{f_{t}}\left\Vert \boldsymbol{f}_{t}-f_{t}\left(\boldsymbol{X},\boldsymbol{t};\Phi,\Theta\right)\right\Vert ^{2}-\log\lambda_{u}\lambda_{f}\lambda_{f_{t}},\nonumber \\
= & \lambda_{u}\mathcal{L}_{u}+\lambda_{f}\mathcal{L}_{f}+\lambda_{f_{t}}\mathcal{L}_{f_{t}}-\log\lambda_{u}\lambda_{f}\lambda_{f_{t}}.\label{eq: Log-Likelihood of multiple tasks}
\end{align}
\end{figure*}
 It can be seen that the loss function (\ref{eq: Loss function})
is regularized as:
\begin{equation}
\mathcal{L}=\lambda_{u}\mathcal{L}_{u}+\lambda_{f}\mathcal{L}_{f}+\lambda_{f_{t}}\mathcal{L}_{f_{t}}-\log\lambda_{u}\lambda_{f}\lambda_{f_{t}}.\label{eq: Regularized loss function}
\end{equation}
With these settings, relative weighting coefficients $\lambda_{u}$,
$\lambda_{f}$, and $\lambda_{f_{t}}$ of the loss terms can be learned
by minimizing the regularized loss function (\ref{eq: Regularized loss function})
adaptively. The output weighted noise is constrained by the penalty
term $\log\lambda_{u}\lambda_{f}\lambda_{f_{t}}$. In practice, we
define $\lambda'\coloneqq-\log\lambda$ and train $\lambda'$ for
numerical stability. The loss function (\ref{eq: Regularized loss function})
can be subsequently rewritten as:
\begin{align}
\mathcal{L} & =\exp\left(-\lambda'_{u}\right)\mathcal{L}_{u}+\exp\left(-\lambda'_{f}\right)\mathcal{L}_{f}+\exp\left(-\lambda'_{f_{t}}\right)\mathcal{L}_{f_{t}}\nonumber \\
 & +\lambda'_{u}+\lambda'_{f}+\lambda'_{f_{t}}.\label{eq: Practical regularized loss function}
\end{align}
Training PINN by minimizing (\ref{eq: Practical regularized loss function})
instead of (\ref{eq: Loss function}) is expected to balance the multiple
losses, which is denoted as AdpBal here. The training process is summarized
in Algorithm 1, where $\mathrm{SSE}\left(\cdot\right)$ represents
computing Sum of Squares Error (SSE) by using eq. (\ref{eq: Loss_u}),
(\ref{eq: Loss_f}), and (\ref{eq: Loss_f_t}), respectively. The
dynamic model $\mathcal{G}\left(\cdot\right)$ can represent the explicit
PDEs or that approximated by the DeepHPM.
\begin{table}
\noindent \centering{}%
\begin{tabular}{>{\raggedright}p{3.3in}}
\toprule 
\multirow{1}{3.3in}{\textbf{\footnotesize{}Algorithm 1}{\footnotesize{} Training proposed
models.}}\tabularnewline
\midrule
\textbf{\footnotesize{}Input:}{\footnotesize{} training data $\mathcal{D}=\left\{ \boldsymbol{X},\boldsymbol{t},\boldsymbol{u}\right\} $,
hyper-parameters.}\tabularnewline
\textbf{\footnotesize{}Output:}{\footnotesize{} surrogate NN $u\left(\boldsymbol{X},\boldsymbol{t};\Phi\right)$,
dynamic model $\mathcal{G}\left(\boldsymbol{X},\boldsymbol{t},\boldsymbol{u},\boldsymbol{u}_{\boldsymbol{x}},\boldsymbol{u}_{\boldsymbol{x}\boldsymbol{x}},\boldsymbol{u}_{\boldsymbol{x}\boldsymbol{x}\boldsymbol{x}},\ldots;\Theta\right)$.}\tabularnewline
{\footnotesize{}\ 1: initialize $\Phi$, $\Theta$, $\lambda'_{u}$,
$\lambda'_{f}$, $\lambda'_{f_{t}}$}\tabularnewline
{\footnotesize{}\ 2: }\textbf{\footnotesize{}for}{\footnotesize{}
$epoch=1,2,\cdots$ }\textbf{\footnotesize{}do:}\tabularnewline
{\footnotesize{}\ 3: \ \ \ \ $\hat{\boldsymbol{u}}\leftarrow u\left(\boldsymbol{X},\boldsymbol{t};\Phi\right)$}\tabularnewline
{\footnotesize{}\ 4: \ \ \ \ $\hat{\boldsymbol{u}}_{t}\leftarrow\mathrm{AutoDiff}\left(\hat{\boldsymbol{u}},\boldsymbol{t}\right)$}\tabularnewline
{\footnotesize{}\ 5: \ \ \ \ $\hat{\boldsymbol{u}}_{\boldsymbol{x}}\leftarrow\mathrm{AutoDiff}\left(\hat{\boldsymbol{u}},\boldsymbol{X}\right)$}\tabularnewline
{\footnotesize{}\ 6: \ \ \ \ $\hat{\boldsymbol{u}}_{\boldsymbol{xx}}\leftarrow\mathrm{AutoDiff}\left(\hat{\boldsymbol{u}}_{\boldsymbol{x}},\boldsymbol{X}\right)$}\tabularnewline
{\footnotesize{}\ 7: \ \ \ \ $\cdots$}\tabularnewline
{\footnotesize{}\ 8: \ \ \ \ $\hat{\mathcal{G}}\leftarrow\mathcal{G}\left(\boldsymbol{X},\boldsymbol{t},\hat{\boldsymbol{u}},\hat{\boldsymbol{u}}_{\boldsymbol{x}},\hat{\boldsymbol{u}}_{\boldsymbol{xx}},\hat{\boldsymbol{u}}_{\boldsymbol{xxx}},\ldots;\Theta\right)$}\tabularnewline
{\footnotesize{}\ 9: \ \ \ \ $\hat{\boldsymbol{f}}\leftarrow\hat{\boldsymbol{u}}_{t}-\hat{\mathcal{G}}$}\tabularnewline
{\footnotesize{}10: \ \ \ \ $\hat{\boldsymbol{f}}_{t}\leftarrow\mathrm{AutoDiff}\left(\hat{\boldsymbol{f}},\boldsymbol{t}\right)$}\tabularnewline
{\footnotesize{}11: \ \ \ \ $\mathcal{\hat{L}}_{u}\leftarrow\mathrm{SSE}\left(\hat{\boldsymbol{u}},\boldsymbol{u}\right)$}\tabularnewline
{\footnotesize{}12: \ \ \ \ $\mathcal{\hat{L}}_{f}\leftarrow\mathrm{SSE}\left(\hat{\boldsymbol{f}},\boldsymbol{0}\right)$}\tabularnewline
{\footnotesize{}13: \ \ \ \ $\mathcal{\hat{L}}_{f_{t}}\leftarrow\mathrm{SSE}\left(\hat{\boldsymbol{f}}_{t},\mathbf{0}\right)$}\tabularnewline
{\footnotesize{}14: \ \ \ \ compute loss $\hat{\mathcal{L}}$
by using eq. (\ref{eq: Practical regularized loss function})}\tabularnewline
{\footnotesize{}15: \ \ \ \ update $\Phi$, $\Theta$, $\lambda'_{u}$,
$\lambda'_{f}$, $\lambda'_{f_{t}}$ on loss $\hat{\mathcal{L}}$}\tabularnewline
{\footnotesize{}16: }\textbf{\footnotesize{}end}\tabularnewline
\bottomrule
\end{tabular}
\end{table}

\section{Dataset Description and Preprocessing\label{sec: Dataset Description and Processing}}

\subsection{Dataset Description}

The dataset involved in this paper consists of three batches of commercial
LFP/graphite battery cells manufactured by A123 Systems, where a total
of 124 cells were cycled to failure under dozens of different fast-charging
protocols \cite{RN385}. The nominal capacity of the cells is 1.1
Ah, and the unit charging and discharging rate 1C equals 1.1 A subsequently.
Each charging protocol is marked as a string formatted as ``C1(Q1)-C2'',
where the corresponding cell was first charged with the current C1
from 0\% SoC (unit: \%) to the SoC Q1. When charged at Q1, the charging
current was then switched to C2 with which the cell was charged to
80\% SoC. All cells were finally charged from 80\% to 100\% SoC with
a Constant-Current Constant-Voltage (CC-CV) form to the 3.6V upper
cutoff potential and the C/50 cutoff current. Moreover, all cells
were discharged also with a CC-CV form at 4C to 2.0V lower cutoff
potential and the C/50 cutoff current. Detailed information on the
studied cells and the experimental settings can be accessed in \cite{RN385}. 

\subsection{Feature Extraction}

\begin{figure}[h]
\centering{}\includegraphics[width=3.1in]{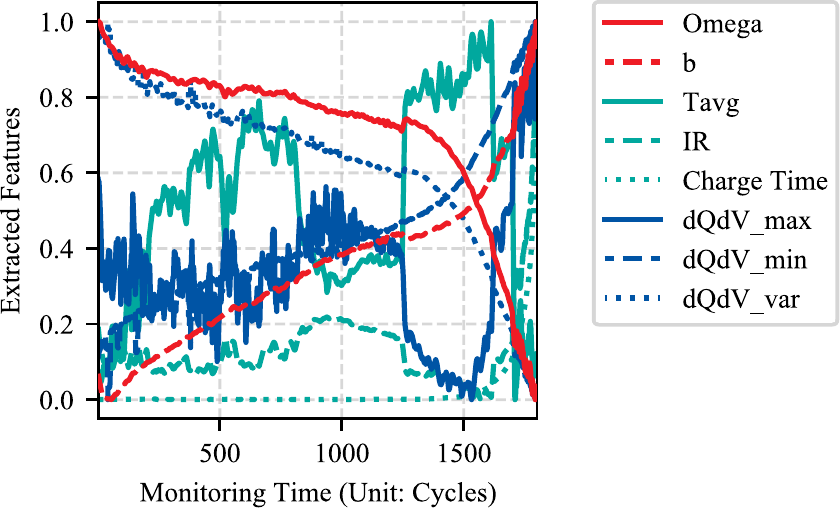}\caption{Extracted features of cell \#124.\label{fig: Features}}
\end{figure}
\begin{figure}[h]
\centering{}\includegraphics[width=3.1in]{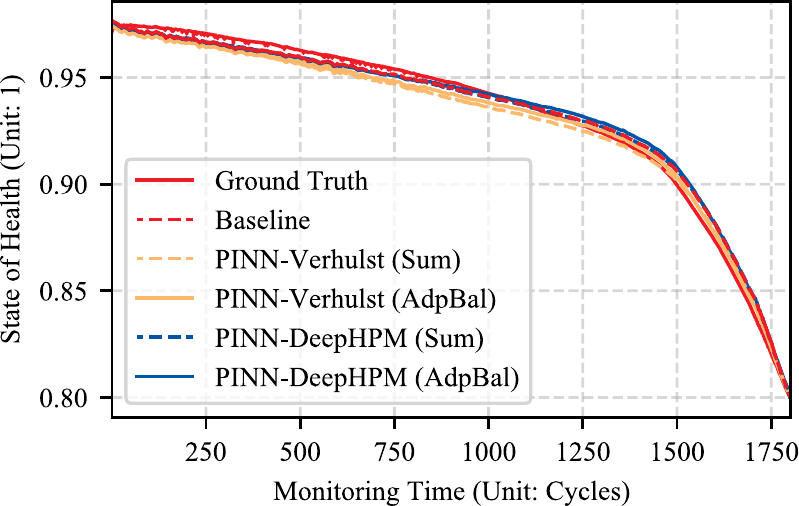}\caption{SoH estimation of cell \#124 in case A.\label{fig: SoH case A}}
\end{figure}
\begin{figure}[h]
\centering{}\includegraphics[width=3.1in]{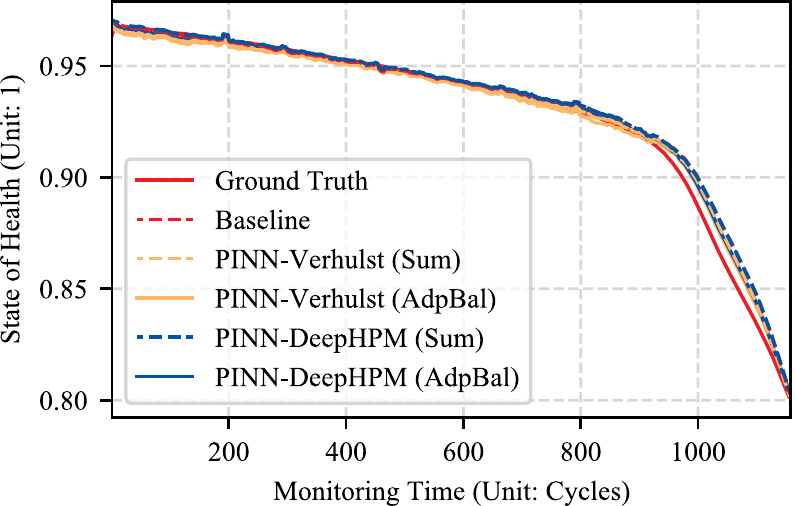}\caption{SoH estimation of cell \#116 in case B.\label{fig: SoH case B}}
\end{figure}
As introduced in Section \ref{sec: Dynamic Models} and \ref{sec: Methodology},
appropriately designed health features $\boldsymbol{x}=\left[x_{1},x_{2},\ldots,x_{S}\right]^{\mathrm{T}}$
can effectively characterize the exact degradation process that each
cell undergoes. Extracting features varying to the number of cycles
for which the battery has operated attracts much attention. A critical
principle is that the extracted features are expected to be of strong
correlation with SoH \cite{RN269}. Point features can be simply extracted
based on the charging/discharging profiles during cycling, such as
the peaks of Incremental Capacity (IC) curves during the CC charging
\cite{RN291}. Point features can be conveniently extracted mainly
characterizing the transient state in each cycle. Interval features
can represent characteristics in a period, and they can also be easily
extracted by intercepting values on the profiles at certain intervals.
Therefore, information related to SoH may be over-simplified since
only values on the endpoints of the intervals are paid attention to.
Interval features can be extracted as the charge time during the CC
charging \cite{RN354} and time intervals of equispaced charge current/voltage
changing \cite{RN342,RN364} for the instance. To capture the trend
variation of charging/discharging profiles during cycling, trend features
are designed focusing on parameters varying upon a specific prior
function modeling the profiles during a cycle. The trend of charging
current during CV charging is approximately exponential, and the parameter
of the fitted exponential function can be used as a trend feature
\cite{RN291}. Within a certain range of discharging voltage, differences
in discharging capacity show a quadratic-function-formed trend to
discharging capacity. Parameters of the quadratic polynomial can be
accordingly estimated as the trend feature as well \cite{RN269}.
Other statistical features computed based on the whole profile of
a cycle can also be employed such as the mean \cite{RN379,RN269},
energy \cite{RN281}, Skewness \cite{RN281}, and Kurtosis \cite{RN281},
etc.

Following \cite{RN269}, we also extracted the trend feature based
on the quadratic model:
\begin{equation}
Q_{i+1}\left(V_{i+1}\right)-Q_{i}\left(V_{i}\right)=-\omega\left[Q_{i}\left(V_{i}\right)\right]^{2}+b+\varepsilon.\label{eq: Quadratic profile}
\end{equation}
Eq. (\ref{eq: Quadratic profile}) is defined for the profile of each
cycle. $Q_{i}\left(V_{i}\right)$ is a measurement of the discharging
capacity with respect to the discharging voltage between 2.7V and
3.3V, and $\varepsilon\sim\mathcal{N}\left(0,\sigma^{2}\right)$ represents
the normal measuring error. Undetermined coefficients $\omega$ and
$b$ are estimated as two features in that cycle. Maximum, minimum,
and the variance of the IC curve when the discharging voltage is between
2.7V and 3.3V are also used as features. Other employed features include
average temperature, internal resistance, and charging time. To reduce
the measurement noise, the moving average is then applied to the extracted
features for better representative capability. All the extracted features
from cell \#124 are illustrated in Fig. \ref{fig: Features} as an
example, and it should be noted that the features are 0-1 normalized
in this figure for better illustration.

\subsection{Standardization}

Note that there are significant variations among different channels
of inputs and outputs, which may severely impact the performance.
Standardization is consequently implemented. The derivatives of outputs
to the inputs can be kept unchanged in this way when using AutoDiff,
and the standardized data are processed by the NN. On the other hand,
the partial derivatives of the before- or after-standardization outputs
to the before- or after-standardization inputs can be obtained. Here
we employ the widely-used z-score standardization to scale inputs.
Specifically, the standardization factors are set as the corresponding
mean and standard deviation calculated from the training data:
\begin{align}
\widetilde{x}_{i} & =\frac{x_{i}-\mathrm{Mean}\left(\boldsymbol{X}_{Train}\right)}{\mathrm{Std}\left(\boldsymbol{X}_{Train}\right)},\label{eq: standardize inputs}\\
\widetilde{u}_{i} & =\frac{u_{i}-\mathrm{Mean}\left(\boldsymbol{u}_{Train}\right)}{\mathrm{Std}\left(\boldsymbol{u}_{Train}\right)}.\label{eq: standardize outputs}
\end{align}
In eq. (\ref{eq: standardize inputs}) and (\ref{eq: standardize outputs}),
$\boldsymbol{u}=\left[u_{1},u_{2},\ldots,u_{N}\right]^{\mathrm{T}}$
and $\boldsymbol{X}=\left[\boldsymbol{x}_{1},\boldsymbol{x}_{2},\ldots,\boldsymbol{x}_{N}\right]^{\mathrm{T}}$
consist of labels and samples of the training set as denoted in Section
\ref{subsec: Weight-Coefficients-Tuning}. Monitoring time $t$ is
also standardized in this way since it can be regarded as part of
inputs.

\section{Case Study\label{sec: Case-Study}}

\begin{table}
\caption{General settings for training NNs\label{tab: General hyper parameters setup to train NN}}

\noindent \centering{}%
\begin{tabular}{ll}
\toprule 
\multirow{1}{*}{{\footnotesize{}Settings}} & {\footnotesize{}Values}\tabularnewline
\midrule
{\footnotesize{}Optimizer} & {\footnotesize{}Adam}\tabularnewline
{\footnotesize{}Training epochs} & {\footnotesize{}2000 (Case A \& B) / 8000 (Case C)}\tabularnewline
{\footnotesize{}Learning rate} & {\footnotesize{}0.001}\tabularnewline
{\footnotesize{}Batch size} & {\footnotesize{}1024 (Case A \& B) / 8192 (Case C)}\tabularnewline
{\footnotesize{}Initialization} & \multicolumn{1}{l}{{\footnotesize{}Xavier Normal}}\tabularnewline
{\footnotesize{}Dropout probability} & {\footnotesize{}0.2}\tabularnewline
\bottomrule
\end{tabular}
\end{table}
The proposed prognostic framework is verified for both SoH estimation
and RUL prediction. For better comparison with the existing methods,
the training set and test set are formed in different ways, which
are marked as cases A, B, and C respectively. In case A, data from
cells \#91 and \#100 are used for training/validation and cell \#124
is used as the test set \cite{RN269}. Their charging protocol is
4.36C(80\%)-4.36C. In case B, data from cells \#101, \#108, and \#120
are used for training/validation and cell \#116 is used as the test
set \cite{RN269}. Their charging protocol follows 5.3C(54\%)-4C.
In case C, data from batch 2 are selected and 20\% of them are randomly
chosen as the test set \cite{RN379}. These cells follow various charging
protocols. 

Here the widely used Root Mean Square Error (RMSE) metric is adopted
to measure the performance. RMSE is formulated as:
\begin{equation}
\mathrm{RMSE}=\sqrt{\frac{1}{N}\sum_{i=1}^{N}\left(\hat{u}_{i}-u_{i}\right)^{2}},\label{eq: RMSE}
\end{equation}
where $\hat{u}_{i}$ and $u_{i}$ denote the output of the prediction
model and the actual data label corresponding to $i^{\mathrm{th}}$
input sample. When it is used for SoH estimation, the output PCL should
be transformed as SoH based on eq. (\ref{eq: PCL and SoH}). There
are a total of $N$ samples in the test set. Moreover, we also adopt
the Root Mean Square Percentage Error (RMSPE) as the metric since
the scale of SoH and RUL can be quite distinct. RMSPE is formulated
as:
\begin{equation}
\mathrm{RMSPE}=\sqrt{\frac{1}{N}\sum_{i=1}^{N}\left(\frac{\hat{u}_{i}-u_{i}}{u_{i}}\right)^{2}}\times100\%.\label{eq: RMSPE}
\end{equation}

\begin{table}
\caption{Optimal settings for each case\label{tab: Summarized hyper parameters of all cases}}

\noindent \centering{}%
\begin{tabular}{llllll}
\toprule 
\multirow{2}{*}{{\footnotesize{}Settings}} & \multicolumn{2}{l}{{\footnotesize{}SoH Estimation}} & \multicolumn{3}{l}{{\footnotesize{}RUL Prognostics}}\tabularnewline
\cmidrule{2-6} \cmidrule{3-6} \cmidrule{4-6} \cmidrule{5-6} \cmidrule{6-6} 
 & {\footnotesize{}Case A} & {\footnotesize{}Case B} & {\footnotesize{}Case A} & {\footnotesize{}Case B} & {\footnotesize{}Case C}\tabularnewline
\midrule
{\footnotesize{}No. of Layers} & {\footnotesize{}2} & {\footnotesize{}2} & {\footnotesize{}2} & {\footnotesize{}2} & {\footnotesize{}4}\tabularnewline
{\footnotesize{}Neurons / Layer} & {\footnotesize{}128} & {\footnotesize{}64} & {\footnotesize{}128} & {\footnotesize{}128} & {\footnotesize{}128}\tabularnewline
{\footnotesize{}DeepHPM inputs} & {\footnotesize{}$\boldsymbol{x},t$} & {\footnotesize{}$t$} & {\footnotesize{}$\boldsymbol{x},t,u$} & {\footnotesize{}$t,u,u_{\boldsymbol{x}}$} & {\footnotesize{}$t$}\tabularnewline
\bottomrule
\end{tabular}
\end{table}
\begin{figure*}
\begin{centering}
\includegraphics[width=6.5in]{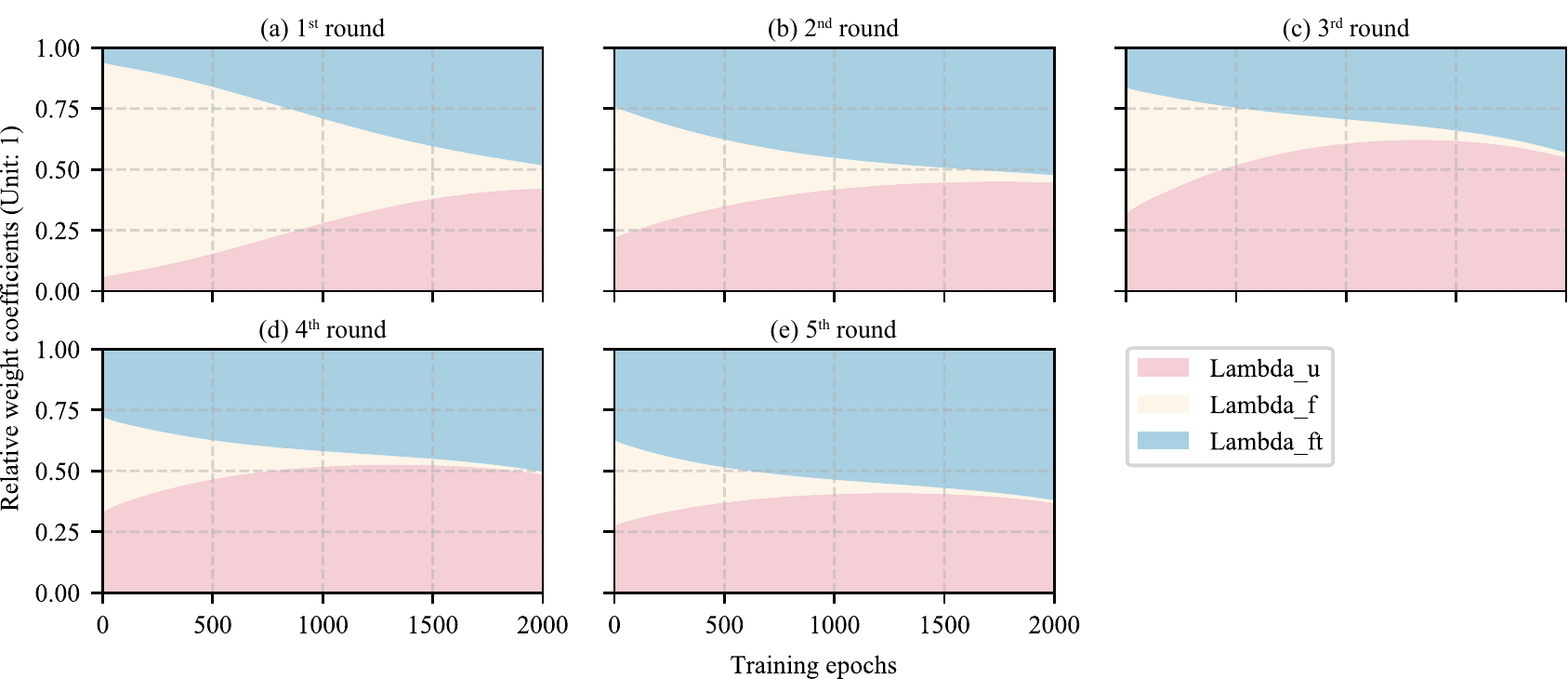}
\par\end{centering}
\caption{Variation of relative weighting coefficients by using adaptive balancing
in the model training. (a)\textasciitilde (e) Respective variation
of relative weighting coefficients in each round.\label{fig: AdpBal coefficients}}
\end{figure*}
Division of training and test data have been detailed above. Then
20\% of training-validation data are divided as the validation data
in cases A and B, and 25\% of training-validation data are divided
as the validation data in case C. The rest of the training-validation
data is used for training ultimately. Several general settings are
listed in Table \ref{tab: General hyper parameters setup to train NN}.
Other critical hyper-parameters such as the number of hidden layers
and neurons are further tuned depending on the performance of validation
data. To reduce the number of hyper-parameters to be tuned, we tune
them based on baseline models as far as possible. Then the obtained
optimal parameters for the baseline models are applied to the proposed
models. It is worth mentioning that those parameters may not be the
optima for the proposed models. For instance, optimal network structures
(the number of hidden layers and neurons) are determined based on
the vanilla data-driven NN illustrated in Fig. \ref{fig: Frameworks-Data driven},
and then they are used for both PINN-Verhulst and PINN-DeepHPM. It
has been revealed that the performance of discovering dynamics by
using DeepHPM can be sensitive to input terms \cite{RN350,RN398,RN399},
so which terms should be involved needs to be further explored. We
organize various combinations of input features, monitoring time as
well as partial derivatives to them as a candidate library. Similarly,
simply summing the multiple losses in eq. (\ref{eq: Loss function})
for training PINNs is further set as a baseline. The optimal combination
of input terms in the library for DeepHPM is determined with this
baseline, and then it is used to validate AdpBal for the multiple
losses. Calculated RMSPEs for SoH estimation and RMSEs for RUL prognostics
on the validation set in terms of the number of hidden layers and
neurons per layer are listed in Tables \ref{tab: Hyper parameters baseline Case A},
\ref{tab: Hyper parameters baseline Case B}, \ref{tab: Hyper parameters baseline RUL Case A},
\ref{tab: Hyper parameters baseline RUL Case B}, and \ref{tab: Hyper parameters baseline RUL Case C}
in the Appendix. Validation RMSPEs and RMSEs in terms of the combinations
of inputs to DeepHPM are listed in Table \ref{tab: DeepHPM inputs}.
Accordingly, settings for different cases are determined and summarized
in Table \ref{tab: Summarized hyper parameters of all cases}. With
these settings, the proposed models are all trained in a regular computation
environment (Intel Xeon E5-2630 CPU, 2.2 GHz; Nvidia GeForce GTX 1080
Ti GPU). The training-test process is repeated for 5 rounds in each
case, and the results are averaged to reduce the randomness.
\begin{table*}
\caption{Summarized results and the comparison for SoH estimation in cases
A and B\label{tab: SoH A and B}}

\noindent \centering{}%
\begin{tabular}{llllllll}
\toprule 
\multirow{2}{*}{{\footnotesize{}Cases}} & \multirow{2}{*}{{\footnotesize{}GPR}} & {\footnotesize{}GPR} & \multirow{2}{*}{{\footnotesize{}Baseline}} & {\footnotesize{}PINN-Verhulst} & {\footnotesize{}PINN-Verhulst} & {\footnotesize{}PINN-DeepHPM} & {\footnotesize{}PINN-DeepHPM}\tabularnewline
 &  & {\footnotesize{}(Onboard)} &  & {\footnotesize{}(Sum)} & {\footnotesize{}(AdpBal)} & {\footnotesize{}(Sum)} & {\footnotesize{}(AdpBal)}\tabularnewline
\midrule
{\footnotesize{}Case A} & {\footnotesize{}0.76} & {\footnotesize{}0.69} & {\footnotesize{}0.42} & {\footnotesize{}1.41} & {\footnotesize{}0.49} & {\footnotesize{}0.43} & {\footnotesize{}0.47}\tabularnewline
\midrule
{\footnotesize{}Case B} & {\footnotesize{}0.63} & {\footnotesize{}0.57} & {\footnotesize{}0.56} & {\footnotesize{}0.56} & {\footnotesize{}0.44} & {\footnotesize{}0.51} & {\footnotesize{}0.42}\tabularnewline
\bottomrule
\end{tabular}
\end{table*}

\subsection{SoH Estimation}

The two categories of dynamic models, the Verhulst model, and DeepHPM
are used to estimate the SoH. This represents two cases that whether
there are available explicit dynamic models respectively. The results
of SoH estimation are summarized in Table \ref{tab: SoH A and B}.

In case A, the calculated RMSPEs of SoH estimation on the validation
data in terms of different combinations of the number of hidden layers
and neurons are listed in Table \ref{tab: Hyper parameters baseline Case A}.
The optimal network structure for the data-driven baseline model is
2 hidden layers and 128 neurons per hidden layer. With this setting,
the baseline model provides a 0.42\% estimation error in terms of
RMSPE on the test data. PINN-Verhulst provides a 1.41\%-RMSPE estimation
when simply summing the multiple losses. With AdpBal for tuning the
weight coefficients of the losses, a 0.49\% RMSPE can be obtained.
When there is no prior explicit dynamic model, the inputs of DeepHPM
are set as {\footnotesize{}$\boldsymbol{x},t$} when discovering the
dynamic model via DeepHPM according to Table \ref{tab: DeepHPM inputs}.
These validation errors are calculated by simply summing the multiple
losses. the estimation errors are 0.43\% and 0.47\% respectively without
and with AdpBal.

In case B, the network structure is set as 2 hidden layers with 64
neurons per hidden layer according to Table \ref{tab: Hyper parameters baseline Case B}.
With this setting, the baseline estimation error is 0.56\%. PINN-Verhulst
(Sum), PINN-Verhulst (AdpBal), PINN-DeepHPM (Sum), and PINN-DeepHPM
(AdpBal) provide 0.56\%, 0.44\%, 0.51\%, and 0.42\% estimation errors
in terms of RMSPEs.

The SoH estimation is further illustrated in Fig. \ref{fig: SoH case A}
and \ref{fig: SoH case B}. The data-driven baseline method without
model fusion can perform satisfactorily for SoH estimation, especially
in case A. With fusing the improved Verhulst model, the performance
of simply summing the losses has no significant improvement over the
baseline. The estimation accuracy is even significantly declined in
case A. By using the AdpBal, the performance is improved instead.
Compared with case A, there is a better improvement in case B. The
reason can be that the numbers of hidden layers and neurons are optimal
for baseline, but probably not for the proposed methods. When DeepHPM
is used to discover the governing dynamics, similar results are observed.
It verifies the ability of DeepHPM to discover dynamics and the robustness
of using AdpBal for training PINN. With AdpBal, relative values of
the weighting coefficients changing in terms of the training epochs
of case B in all 5 rounds are illustrated in Fig. \ref{fig: AdpBal coefficients}.
It can be seen that relative values of the weighting coefficients
vary in similar ways during different rounds of training. This also
shows the robustness of AdpBal. The average training time of the baseline
data-driven model, PINN-Verhulst, and PINN-DeepHPM in case B are 87.6s,
122.7s, and 126.0s respectively. Moreover, the proposed methods are
compared with a Gaussian Process Regression (GPR)-based method \cite{RN269}.
It is worth mentioning that partial onboard test data are further
used to calibrate this GPR-based method (GPR Onboard) in \cite{RN269}.
It should be noted that the results of the methods compared here are
directly quoted from the corresponding papers. This comparison is
shown in Table \ref{tab: SoH A and B}. As can be seen, the proposed
methods perform better than this comparative study even without using
any onboard test data.
\begin{figure}[h]
\centering{}\includegraphics[width=3.1in]{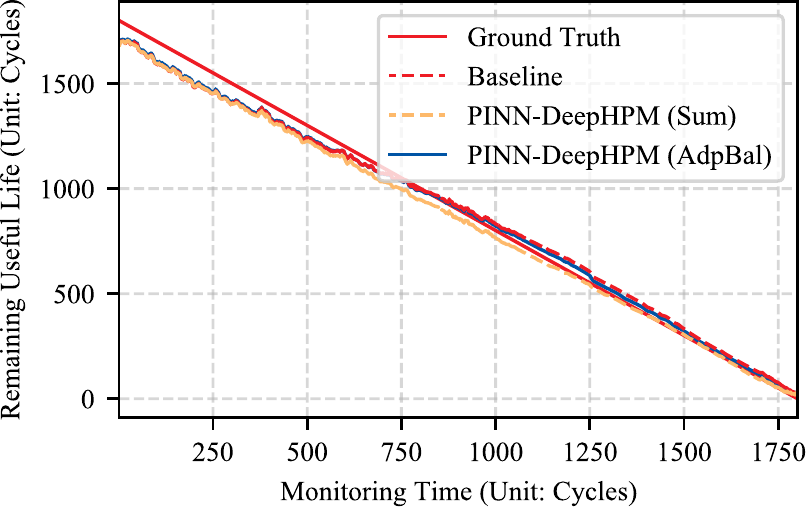}\caption{RUL prognostics of cell \#124 in case A.\label{fig: RUL case A}}
\end{figure}
\begin{figure}[h]
\centering{}\includegraphics[width=3.1in]{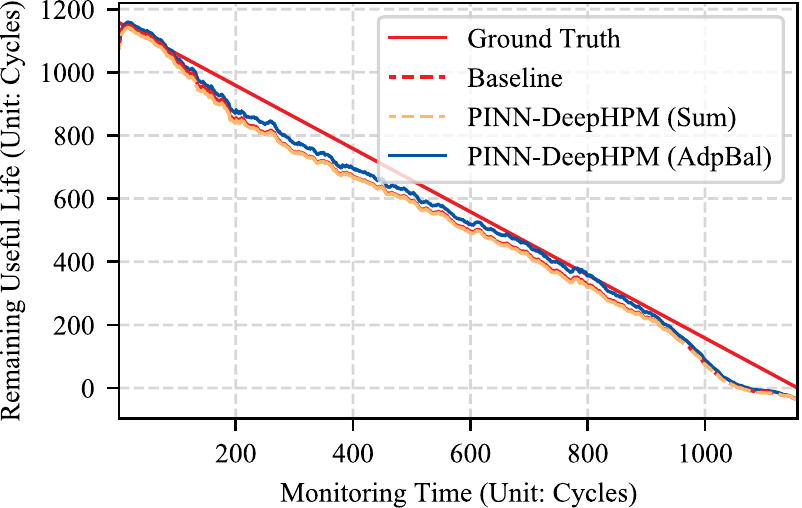}\caption{RUL prognostics of cell \#116 in case B.\label{fig: RUL case B}}
\end{figure}

\subsection{RUL Prognostics}

\begin{table}[h]
\caption{Summarized results and the comparison for RUL prognostics in cases
A and B\label{tab: RUL A and B}}

\noindent \centering{}%
\begin{tabular}{lllll}
\toprule 
\multirow{2}{*}{{\footnotesize{}Cases}} & \multirow{2}{*}{{\footnotesize{}GPR}} & \multirow{2}{*}{{\footnotesize{}Baseline}} & {\footnotesize{}PINN-DeepHPM} & {\footnotesize{}PINN-DeepHPM}\tabularnewline
 &  &  & {\footnotesize{}(Sum)} & {\footnotesize{}(AdpBal)}\tabularnewline
\midrule
{\footnotesize{}Case A} & {\footnotesize{}93.27} & {\footnotesize{}46.38} & {\footnotesize{}48.81} & {\footnotesize{}45.86}\tabularnewline
\midrule
{\footnotesize{}Case B} & {\footnotesize{}64.06} & {\footnotesize{}64.48} & {\footnotesize{}65.52} & {\footnotesize{}56.29}\tabularnewline
\bottomrule
\end{tabular}
\end{table}
\begin{table*}
\caption{Summarized results and the comparison for RUL prognostics in case
C\label{tab: RUL C}}

\noindent \centering{}%
\begin{tabular}{lllllllll}
\toprule 
\multirow{2}{*}{{\footnotesize{}Cases}} & \multirow{2}{*}{{\footnotesize{}LSTM}} & {\footnotesize{}Heteroscedastic} & {\footnotesize{}Homoscedastic} & {\footnotesize{}Bayesian} & {\footnotesize{}Bayesian DL} & \multirow{2}{*}{{\footnotesize{}Baseline}} & {\footnotesize{}PINN-DeepHPM} & {\footnotesize{}PINN-DeepHPM}\tabularnewline
 &  & {\footnotesize{}Bayesian LSTM} & {\footnotesize{}Bayesian LSTM} & {\footnotesize{}VAE-LSTM} & {\footnotesize{}(Calibration)} &  & {\footnotesize{}(Sum)} & {\footnotesize{}(AdpBal)}\tabularnewline
\midrule
{\footnotesize{}Case C} & {\footnotesize{}15.8} & {\footnotesize{}20.7} & {\footnotesize{}22.4} & {\footnotesize{}22.4} & {\footnotesize{}15.2} & {\footnotesize{}15.2} & {\footnotesize{}14.2} & {\footnotesize{}17.9}\tabularnewline
\bottomrule
\end{tabular}
\end{table*}
Since there is no prior explicit dynamic PDE model correlating the
predicted RUL to the input features as well as monitoring time, only
DeepHPM is employed here to discover the dynamics. The results are
summarized in Tables \ref{tab: RUL A and B} and \ref{tab: RUL C}.

Predicted and actual RUL of the test cells \#124 and \#116 are illustrated
in Fig. \ref{fig: RUL case A} and \ref{fig: RUL case B}. In case
A, the data-driven prognostic model provides a 46.38 cycles error
in terms of RMSE. With physics-informed losses based on the discovery
of DeepHPM, the prediction error is 48.81 cycles. After AdpBal is
turned on, the predictive RMSE is 45.86 cycles. The prognostic error
is slightly reduced by 6.04\% ((45.86$-$48.81)$\text{/}$48.81$\times$100\%)
after weighting coefficients are tuned automatically. In case B, the
prediction RMSE of baseline, PINN-DeepHPM (Sum), and PINN-DeepHPM
(AdpBal) are 64.48, 65.52, and 56.29 cycles respectively. The prognostic
error is significantly reduced by 14.09\% ((56.29$-$65.52)$\text{/}$65.52$\times$100\%)
after weighting coefficients are tuned automatically. Similar to that
of SoH estimation, the proposed method shows a minor improvement in
case A and a major improvement in case B. As a result, the generalizability
of the proposed methods from SoH estimation to RUL prognostics is
verified. In case C, RUL prognostics error is reduced by 6.7\% ((14.19$-$15.21)$\text{/}$15.21$\times$100\%)
with introducing DeepHPM. However, the performance deteriorated by
using AdpBal. A possible reason can be that there are multiple operation
conditions for the cells in case C, which severely impacts the homoscedastic
uncertainty among measurements of cells. Moreover, the proposed methods
are compared with an LSTM NN, a Heteroscedastic Bayesian LSTM, a Homoscedastic
Bayesian LSTM, a Bayesian VAE-LSTM, and a Bayesian DL framework incorporating
uncertainty quantification and calibration \cite{RN379}. It can be
seen from Table \ref{tab: RUL C} that the fused PINN-DeepHPM outperforms
those LSTM-based methods. LSTM commonly fits temporal information
better than vanilla NN since it can capture the dynamics by using
differential equations. The fusion with DeepHPM enables the vanilla
NN to capture the dynamics as well.

\subsection{Discussion}

In the PHM of Li-ion batteries, electrochemical models with higher
complexity may provide better estimations, such as the SP models.
However, the methods to specify the model parameters need further
development since the automatic discovery of parameters can be challenging
in that case. In this paper, only the vanilla NN is used to construct
the PINN. The performance may be seriously limited due to the simple
structure of the vanilla NN. On the other hand, there is still a lack
of research on how more powerful architectures of NNs, such as the
RNN, the Convolutional Neural Network (CNN), and their variants, will
affect the propagation and differentiation in PINN. For example, these
advanced network structures can relate a single output $u_{k}$ to
multiple inputs $\left(\boldsymbol{x}_{i},t_{i}\right),\,i=1,2,\cdots,N$
in the time domain. This modification will profoundly affect the dynamic
model (\ref{eq: Basic PDE}) since all partial differentials will
vary with the monitoring time in this equation. As a result, more
theoretical research needs to be conducted to fill this critical gap.

\section{Conclusion\label{sec:Conclusion}}

In this article, we propose a new model fusion framework for Li-ion
batteries PHM based on PINN. A flexible form is provided by PINN to
fuse empirical or physical dynamic models and data-driven surrogate
NNs. With the established dynamic model, the rate of capacity degrading
can be quantified with respect to the operating cycles and the current
SoH of monitored cells. The generalization of the dynamic model to
a PDE form fits different cells with the same combination of parameters.
The added extra parameter can represent the initial SEI formation,
which is a common reason for capacity loss in the initial cycles.
Under this setting, the model is more consistent with the physical
characteristics of battery degradation. DeepHPM provides a specialized
model to discover the governing dynamics of battery degradation when
lacking prior information. With the uncertainty-based weighting method,
losses of multiple learning tasks can be adaptively balanced when
training the PINN. Implementation of a public dataset verifies the
effectiveness of the proposed methods. The results show that the proposed
model fusion scheme can improve the performance of PHM for Li-ion
batteries, but an appropriate weighting method is compulsory to balance
the multi-task losses for training PINN.

In future directions, special attention should be paid to the issues
of possible failure to learn complex physical phenomena for general
implementations of PINN. To enhance the performance of PINN, more
efforts can be spent on improving the convergence, stability, boundary
conditions, network design, and optimization \cite{RN478}. How to
propagate the gradients along the temporal dimension in RNN or spatial
dimension in CNN and their variants can be quite a potential direction.
Another possible direction together with this issue is how to integrate
the temporal or spatial gradients into the PDEs.

\appendices{}

\section{Performance of Different Settings on the Validation Data\label{sec: Hyper-Parameters-Tuning}}

\subsection{NN Structures for SoH Estimation\label{subsec: Tuning SoH NN structures}}

In terms of the different number of hidden layers and neurons per
layer, the calculated RMSPEs for SoH estimation on the validation
set in respective cases are listed in Tables \ref{tab: Hyper parameters baseline Case A}
and \ref{tab: Hyper parameters baseline Case B}. These results are
calculated by using the vanilla NN (Baseline).
\begin{table}[H]
\noindent \begin{centering}
\caption{Validation RMSPE with respect to the number of hidden layers and neurons
per layer in Case A based on the baseline model (Unit: \%)\label{tab: Hyper parameters baseline Case A}}
\par\end{centering}
\noindent \centering{}%
\begin{tabular}{>{\raggedright}p{1.5cm}lllll}
\toprule 
{\footnotesize{}\diagbox[width=2cm]{Layers}{Neurons}} & {\footnotesize{}8} & {\footnotesize{}16} & {\footnotesize{}32} & {\footnotesize{}64} & {\footnotesize{}128}\tabularnewline
\midrule 
{\footnotesize{}2} & {\footnotesize{}2.4e-01} & {\footnotesize{}1.4e-01} & {\footnotesize{}1.0e-01} & {\footnotesize{}8.9e-02} & \textbf{\footnotesize{}7.9e-02}\tabularnewline
{\footnotesize{}4} & {\footnotesize{}3.7e-01} & {\footnotesize{}2.2e-01} & {\footnotesize{}1.5e-01} & {\footnotesize{}1.2e-01} & {\footnotesize{}1.5e-01}\tabularnewline
{\footnotesize{}6} & {\footnotesize{}5.0e-01} & {\footnotesize{}2.7e-01} & {\footnotesize{}1.9e-01} & {\footnotesize{}1.6e-01} & {\footnotesize{}2.0e-01}\tabularnewline
{\footnotesize{}8} & {\footnotesize{}6.4e-01} & {\footnotesize{}3.3e-01} & {\footnotesize{}2.2e-01} & {\footnotesize{}2.0e-01} & {\footnotesize{}2.0e-01}\tabularnewline
{\footnotesize{}10} & {\footnotesize{}8.3e-01} & {\footnotesize{}5.1e-01} & {\footnotesize{}4.1e-01} & {\footnotesize{}2.7e-01} & {\footnotesize{}2.4e-01}\tabularnewline
\bottomrule
\end{tabular}
\end{table}
\begin{table}[H]
\caption{Validation RMSPE with respect to the number of hidden layers and neurons
per layer in Case B based on the baseline model (Unit: \%)\label{tab: Hyper parameters baseline Case B}}

\noindent \centering{}%
\begin{tabular}{>{\raggedright}p{1.5cm}lllll}
\toprule 
{\footnotesize{}\diagbox[width=2cm]{Layers}{Neurons}} & {\footnotesize{}8} & {\footnotesize{}16} & {\footnotesize{}32} & {\footnotesize{}64} & {\footnotesize{}128}\tabularnewline
\midrule 
{\footnotesize{}2} & {\footnotesize{}2.7e-01} & {\footnotesize{}1.8e-01} & {\footnotesize{}1.4e-01} & \textbf{\footnotesize{}1.1e-01} & {\footnotesize{}1.1e-01}\tabularnewline
{\footnotesize{}4} & {\footnotesize{}4.0e-01} & {\footnotesize{}2.5e-01} & {\footnotesize{}1.8e-01} & {\footnotesize{}1.7e-01} & {\footnotesize{}1.7e-01}\tabularnewline
{\footnotesize{}6} & {\footnotesize{}4.4e-01} & {\footnotesize{}3.2e-01} & {\footnotesize{}2.3e-01} & {\footnotesize{}1.9e-01} & {\footnotesize{}2.2e-01}\tabularnewline
{\footnotesize{}8} & {\footnotesize{}5.1e-01} & {\footnotesize{}3.5e-01} & {\footnotesize{}2.5e-01} & {\footnotesize{}2.3e-01} & {\footnotesize{}2.3e-01}\tabularnewline
{\footnotesize{}10} & {\footnotesize{}6.2e-01} & {\footnotesize{}5.4e-01} & {\footnotesize{}3.8e-01} & {\footnotesize{}2.8e-01} & {\footnotesize{}2.8e-01}\tabularnewline
\bottomrule
\end{tabular}
\end{table}

\subsection{NN Structures for RUL Prognostics\label{subsec: Tuning RUL NN structures}}

In terms of the different number of hidden layers and neurons per
layer, calculated RMSEs for RUL prognostics on the validation set
in respective cases are listed in Tables \ref{tab: Hyper parameters baseline RUL Case A},
\ref{tab: Hyper parameters baseline RUL Case B}, and \ref{tab: Hyper parameters baseline RUL Case C}.
These results are calculated by using the vanilla NN (Baseline).
\begin{table}[H]
\caption{Validation RMSE with respect to the number of hidden layers and neurons
per layer in Case A based on the baseline model (Unit: Cycles)\label{tab: Hyper parameters baseline RUL Case A}}

\noindent \centering{}%
\begin{tabular}{>{\raggedright}p{1.5cm}lllll}
\toprule 
{\footnotesize{}\diagbox[width=2cm]{Layers}{Neurons}} & {\footnotesize{}8} & {\footnotesize{}16} & {\footnotesize{}32} & {\footnotesize{}64} & {\footnotesize{}128}\tabularnewline
\midrule 
{\footnotesize{}2} & {\footnotesize{}29.45} & {\footnotesize{}19.82} & {\footnotesize{}13.26} & {\footnotesize{}10.39} & \textbf{\footnotesize{}8.31}\tabularnewline
{\footnotesize{}4} & {\footnotesize{}50.32} & {\footnotesize{}29.55} & {\footnotesize{}20.65} & {\footnotesize{}15.50} & {\footnotesize{}15.32}\tabularnewline
{\footnotesize{}6} & {\footnotesize{}58.56} & {\footnotesize{}42.38} & {\footnotesize{}24.00} & {\footnotesize{}21.43} & {\footnotesize{}21.65}\tabularnewline
{\footnotesize{}8} & {\footnotesize{}63.49} & {\footnotesize{}48.25} & {\footnotesize{}29.05} & {\footnotesize{}24.96} & {\footnotesize{}22.06}\tabularnewline
{\footnotesize{}10} & {\footnotesize{}68.40} & {\footnotesize{}51.97} & {\footnotesize{}38.42} & {\footnotesize{}26.44} & {\footnotesize{}25.57}\tabularnewline
\bottomrule
\end{tabular}
\end{table}
\begin{table}[H]
\caption{Validation RMSE with respect to the number of hidden layers and neurons
per layer in Case B based on the baseline model (Unit: Cycles)\label{tab: Hyper parameters baseline RUL Case B}}

\noindent \centering{}%
\begin{tabular}{>{\raggedright}p{1.5cm}lllll}
\toprule 
{\footnotesize{}\diagbox[width=2cm]{Layers}{Neurons}} & {\footnotesize{}8} & {\footnotesize{}16} & {\footnotesize{}32} & {\footnotesize{}64} & {\footnotesize{}128}\tabularnewline
\midrule 
{\footnotesize{}2} & {\footnotesize{}35.75} & {\footnotesize{}23.67} & {\footnotesize{}17.12} & {\footnotesize{}14.81} & \textbf{\footnotesize{}12.73}\tabularnewline
{\footnotesize{}4} & {\footnotesize{}44.52} & {\footnotesize{}27.59} & {\footnotesize{}20.27} & {\footnotesize{}15.31} & {\footnotesize{}14.55}\tabularnewline
{\footnotesize{}6} & {\footnotesize{}50.79} & {\footnotesize{}30.14} & {\footnotesize{}24.22} & {\footnotesize{}18.06} & {\footnotesize{}17.40}\tabularnewline
{\footnotesize{}8} & {\footnotesize{}56.89} & {\footnotesize{}41.24} & {\footnotesize{}28.29} & {\footnotesize{}20.41} & {\footnotesize{}16.63}\tabularnewline
{\footnotesize{}10} & {\footnotesize{}62.02} & {\footnotesize{}40.63} & {\footnotesize{}34.08} & {\footnotesize{}21.15} & {\footnotesize{}17.19}\tabularnewline
\bottomrule
\end{tabular}
\end{table}
\begin{table}[H]
\caption{Validation RMSE with respect to the number of hidden layers and neurons
per layer in Case C based on the baseline model (Unit: Cycles)\label{tab: Hyper parameters baseline RUL Case C}}

\noindent \centering{}%
\begin{tabular}{>{\raggedright}p{1.5cm}lllll}
\toprule 
{\footnotesize{}\diagbox[width=2cm]{Layers}{Neurons}} & {\footnotesize{}8} & {\footnotesize{}16} & {\footnotesize{}32} & {\footnotesize{}64} & {\footnotesize{}128}\tabularnewline
\midrule 
{\footnotesize{}2} & {\footnotesize{}107.02} & {\footnotesize{}77.33} & {\footnotesize{}53.88} & {\footnotesize{}35.50} & {\footnotesize{}23.38}\tabularnewline
{\footnotesize{}4} & {\footnotesize{}113.65} & {\footnotesize{}77.40} & {\footnotesize{}46.75} & {\footnotesize{}23.31} & \textbf{\footnotesize{}14.92}\tabularnewline
{\footnotesize{}6} & {\footnotesize{}119.27} & {\footnotesize{}81.05} & {\footnotesize{}48.38} & {\footnotesize{}23.99} & {\footnotesize{}15.02}\tabularnewline
{\footnotesize{}8} & {\footnotesize{}129.26} & {\footnotesize{}87.30} & {\footnotesize{}54.36} & {\footnotesize{}27.59} & {\footnotesize{}15.26}\tabularnewline
{\footnotesize{}10} & {\footnotesize{}144.46} & {\footnotesize{}95.78} & {\footnotesize{}58.84} & {\footnotesize{}29.97} & {\footnotesize{}17.00}\tabularnewline
\bottomrule
\end{tabular}
\end{table}
\begin{table}[H]
\caption{Validation metrics with respect to the combinations of inputs of DeepHPM
based on the PINN-DeepHPM with simply summing losses (Unit: \% for
SoH estimation; Cycles for RUL prognostics)\label{tab: DeepHPM inputs}}

\noindent \centering{}%
\begin{tabular}{>{\raggedright}p{0.5in}lllll}
\toprule 
\multirow{2}{0.5in}{{\footnotesize{}DeepHPM Inputs}} & \multicolumn{2}{l}{{\footnotesize{}SoH Estimation}} & \multicolumn{3}{l}{{\footnotesize{}RUL Prognostics}}\tabularnewline
\cmidrule{2-6} \cmidrule{3-6} \cmidrule{4-6} \cmidrule{5-6} \cmidrule{6-6} 
 & {\footnotesize{}Case A} & {\footnotesize{}Case B} & {\footnotesize{}Case A} & {\footnotesize{}Case B} & {\footnotesize{}Case C}\tabularnewline
\midrule 
{\footnotesize{}$\boldsymbol{x}$} & {\footnotesize{}9.7e-02} & {\footnotesize{}1.1e-01} & {\footnotesize{}9.17} & {\footnotesize{}11.70} & {\footnotesize{}15.52}\tabularnewline
{\footnotesize{}$t$} & {\footnotesize{}9.7e-02} & \textbf{\footnotesize{}1.1e-01} & {\footnotesize{}9.13} & {\footnotesize{}12.02} & \textbf{\footnotesize{}14.05}\tabularnewline
{\footnotesize{}$u$} & {\footnotesize{}1.6e-01} & {\footnotesize{}2.7e-01} & {\footnotesize{}9.13} & {\footnotesize{}12.02} & {\footnotesize{}14.07}\tabularnewline
{\footnotesize{}$u_{\boldsymbol{x}}$} & {\footnotesize{}2.6e-01} & {\footnotesize{}2.9e-01} & {\footnotesize{}9.17} & {\footnotesize{}11.70} & {\footnotesize{}26.00}\tabularnewline
{\footnotesize{}$\boldsymbol{x},t$} & \textbf{\footnotesize{}7.9e-02} & {\footnotesize{}1.1e-01} & {\footnotesize{}8.76} & {\footnotesize{}12.16} & {\footnotesize{}15.04}\tabularnewline
{\footnotesize{}$\boldsymbol{x},u$} & {\footnotesize{}1.9e-01} & {\footnotesize{}2.1e-01} & {\footnotesize{}8.76} & {\footnotesize{}12.16} & {\footnotesize{}15.04}\tabularnewline
{\footnotesize{}$\boldsymbol{x},u_{\boldsymbol{x}}$} & {\footnotesize{}2.6e-01} & {\footnotesize{}3.2e-01} & {\footnotesize{}8.80} & {\footnotesize{}12.38} & {\footnotesize{}23.35}\tabularnewline
{\footnotesize{}$t,u$} & {\footnotesize{}2.2e-01} & {\footnotesize{}1.9e-01} & {\footnotesize{}9.11} & {\footnotesize{}11.89} & {\footnotesize{}15.18}\tabularnewline
{\footnotesize{}$t,u_{\boldsymbol{x}}$} & {\footnotesize{}2.5e-01} & {\footnotesize{}2.7e-01} & {\footnotesize{}8.75} & {\footnotesize{}12.16} & {\footnotesize{}26.59}\tabularnewline
{\footnotesize{}$u,u_{\boldsymbol{x}}$} & {\footnotesize{}2.6e-01} & {\footnotesize{}3.2e-01} & {\footnotesize{}8.76} & {\footnotesize{}12.17} & {\footnotesize{}28.29}\tabularnewline
{\footnotesize{}$\boldsymbol{x},t,u$} & {\footnotesize{}1.8e-01} & {\footnotesize{}2.1e-01} & \textbf{\footnotesize{}8.40} & {\footnotesize{}11.69} & {\footnotesize{}15.90}\tabularnewline
{\footnotesize{}$\boldsymbol{x},t,u_{\boldsymbol{x}}$} & {\footnotesize{}2.5e-01} & {\footnotesize{}2.8e-01} & {\footnotesize{}8.97} & {\footnotesize{}12.36} & {\footnotesize{}21.99}\tabularnewline
{\footnotesize{}$\boldsymbol{x},u,u_{\boldsymbol{x}}$} & {\footnotesize{}2.7e-01} & {\footnotesize{}2.9e-01} & {\footnotesize{}8.96} & {\footnotesize{}12.34} & {\footnotesize{}24.40}\tabularnewline
{\footnotesize{}$t,u,u_{\boldsymbol{x}}$} & {\footnotesize{}3.0e-01} & {\footnotesize{}2.9e-01} & {\footnotesize{}8.41} & \textbf{\footnotesize{}11.68} & {\footnotesize{}22.77}\tabularnewline
{\footnotesize{}$\boldsymbol{x},t,u,u_{\boldsymbol{x}}$} & {\footnotesize{}2.8e-01} & {\footnotesize{}2.7e-01} & {\footnotesize{}9.31} & {\footnotesize{}11.71} & {\footnotesize{}24.98}\tabularnewline
\bottomrule
\end{tabular}
\end{table}

\subsection{Combinations of Inputs of DeepHPM}

In terms of the different combinations of inputs to the DeepHPM, calculated
metrics on the validation set in respective cases are listed in Table
\ref{tab: DeepHPM inputs}. With the NN structures that are determined
according to Appendices \ref{subsec: Tuning SoH NN structures} and
\ref{subsec: Tuning RUL NN structures}, these results are calculated
by using the PINN-DeepHPM by simply summing the losses (Sum).

\bibliographystyle{IEEEtran}
\bibliography{PINN_Prognostics}

\begin{thebibliography}{10}
\providecommand{\url}[1]{#1}
\csname url@samestyle\endcsname
\providecommand{\newblock}{\relax}
\providecommand{\bibinfo}[2]{#2}
\providecommand{\BIBentrySTDinterwordspacing}{\spaceskip=0pt\relax}
\providecommand{\BIBentryALTinterwordstretchfactor}{4}
\providecommand{\BIBentryALTinterwordspacing}{\spaceskip=\fontdimen2\font plus
\BIBentryALTinterwordstretchfactor\fontdimen3\font minus
  \fontdimen4\font\relax}
\providecommand{\BIBforeignlanguage}[2]{{%
\expandafter\ifx\csname l@#1\endcsname\relax
\typeout{** WARNING: IEEEtran.bst: No hyphenation pattern has been}%
\typeout{** loaded for the language `#1'. Using the pattern for}%
\typeout{** the default language instead.}%
\else
\language=\csname l@#1\endcsname
\fi
#2}}
\providecommand{\BIBdecl}{\relax}
\BIBdecl

\bibitem{RN405}
E.~M. Bibra, E.~Connelly, S.~Dhir, M.~Drtil, P.~Henriot, I.~Hwang, J.-B.
  Le~Marois, S.~McBain, L.~Paoli, and J.~Teter, ``Global {EV} outlook 2022:
  Securing supplies for an electric future,'' Int. Energy Agency, Report, 2022.

\bibitem{RN475}
Z.~Wei, Y.~Li, and L.~Cai, ``Electric vehicle charging scheme for a
  park-and-charge system considering battery degradation costs,'' \emph{{IEEE}
  Trans. Intell. Veh.}, vol.~3, no.~3, pp. 361--373, 2018.

\bibitem{RN477}
S.~Schaut, E.~Arnold, and O.~Sawodny, ``Predictive thermal management for an
  electric vehicle powertrain,'' \emph{{IEEE} Trans. Intell. Veh.}, vol.~8,
  no.~2, pp. 1957--1970, 2023.

\bibitem{RN141}
L.~X. Liao and F.~Kottig, ``Review of hybrid prognostics approaches for
  remaining useful life prediction of engineered systems, and an application to
  battery life prediction,'' \emph{{IEEE} Trans. Rel.}, vol.~63, no.~1, pp.
  191--207, 2014.

\bibitem{RN408}
S.~Zhao, S.~Chen, F.~Yang, E.~Ugur, B.~Akin, and H.~Wang, ``A composite failure
  precursor for condition monitoring and remaining useful life prediction of
  discrete power devices,'' \emph{{IEEE} Trans. Ind. Informat.}, vol.~17,
  no.~1, pp. 688--698, 2020.

\bibitem{RN476}
K.~Sarrafan, K.~M. Muttaqi, and D.~Sutanto, ``Real-time state-of-charge
  tracking embedded in the advanced driver assistance system of electric
  vehicles,'' \emph{{IEEE} Trans. Intell. Veh.}, vol.~5, no.~3, pp. 497--507,
  2020.

\bibitem{RN204}
M.~Pecht, T.~Shibutani, M.~Kang, M.~Hodkiewicz, and E.~Cripps, ``A fusion
  prognostics-based qualification test methodology for microelectronic
  products,'' \emph{Microelectron. Rel.}, vol.~63, pp. 320--324, 2016.

\bibitem{RN205}
L.~Liao and F.~Koettig, ``A hybrid framework combining data-driven and
  model-based methods for system remaining useful life prediction,''
  \emph{Appl. Soft Comput.}, vol.~44, pp. 191--199, 2016.

\bibitem{RN130}
Y.~Z. Zhang, R.~Xiong, H.~W. He, and M.~Pecht, ``Validation and verification of
  a hybrid method for remaining useful life prediction of lithium-ion
  batteries,'' \emph{J. Cleaner Prod.}, vol. 212, pp. 240--249, 2019.

\bibitem{RN362}
G.~E. Karniadakis, I.~G. Kevrekidis, L.~Lu, P.~Perdikaris, S.~F. Wang, and
  L.~Yang, ``Physics-informed machine learning,'' \emph{Nat. Rev. Phys.},
  vol.~3, no.~6, pp. 422--440, 2021.

\bibitem{RN287}
S.~Shaier, M.~Raissi, and P.~Seshaiyer, ``Data-driven approaches for predicting
  spread of infectious diseases through {DINN}s: Disease informed neural
  networks,'' \emph{Preprint arXiv:2110.05445v2}, 2021.

\bibitem{RN400}
S.~Zhao, Y.~Peng, Y.~Zhang, and H.~Wang, ``Parameter estimation of power
  electronic converters with physics-informed machine learning,'' \emph{{IEEE}
  Trans. Power Electron.}, vol.~37, no.~10, pp. 11\,567--11\,578, 2022.

\bibitem{RN401}
S.~Zhao, F.~Blaabjerg, and H.~Wang, ``An overview of artificial intelligence
  applications for power electronics,'' \emph{{IEEE} Trans. Power Electron.},
  vol.~36, no.~4, pp. 4633--4658, 2020.

\bibitem{RN406}
H.~Y. Sun, L.~S. Peng, J.~M. Lin, S.~Wang, W.~Zhao, and S.~L. Huang,
  ``Microcrack defect quantification using a focusing high-order {SH} guided
  wave {EMAT}: The physics-informed deep neural network {GuwNet},''
  \emph{{IEEE} Trans. Ind. Informat.}, vol.~18, no.~5, pp. 3235--3247, 2022.

\bibitem{RN407}
H.~Y. Sun, L.~S. Peng, S.~L. Huang, S.~S. Li, Y.~Long, S.~Wang, and W.~Zhao,
  ``Development of a physics-informed doubly fed cross-residual deep neural
  network for high-precision magnetic flux leakage defect size estimation,''
  \emph{{IEEE} Trans. Ind. Informat.}, vol.~18, no.~3, pp. 1629--1640, 2022.

\bibitem{RN278}
W.~Xian, B.~Long, M.~Li, and H.~Wang, ``Prognostics of lithium-ion batteries
  based on the verhulst model, particle swarm optimization and particle
  filter,'' \emph{{IEEE} Trans. Instrum. Meas.}, vol.~63, no.~1, pp. 2--17,
  2013.

\bibitem{RN403}
A.~Jokar, B.~Rajabloo, M.~D\'esilets, and M.~Lacroix, ``Review of simplified
  pseudo-two-dimensional models of lithium-ion batteries,'' \emph{J. Power
  Sour.}, vol. 327, pp. 44--55, 2016.

\bibitem{RN282}
J.~Li, K.~Adewuyi, N.~Lotfi, R.~G. Landers, and J.~Park, ``A single particle
  model with chemical/mechanical degradation physics for lithium ion battery
  {State of Health (SOH)} estimation,'' \emph{Appl. Energy}, vol. 212, pp.
  1178--1190, 2018.

\bibitem{RN285}
R.~Deshpande, M.~Verbrugge, Y.-T. Cheng, J.~Wang, and P.~Liu, ``Battery cycle
  life prediction with coupled chemical degradation and fatigue mechanics,''
  \emph{J. Electrochem. Soc.}, vol. 159, no.~10, p. A1730, 2012.

\bibitem{RN404}
J.-Z. Kong, D.~Wang, T.~Yan, J.~Zhu, and X.~Zhang, ``Accelerated stress factors
  based nonlinear wiener process model for lithium-ion battery prognostics,''
  \emph{{IEEE} Trans. Ind. Electron.}, vol.~69, no.~11, pp. 11\,665--11\,674,
  2021.

\bibitem{RN277}
B.~Xu, A.~Oudalov, A.~Ulbig, G.~Andersson, and D.~S. Kirschen, ``Modeling of
  lithium-ion battery degradation for cell life assessment,'' \emph{{IEEE}
  Trans. Smart Grid}, vol.~9, no.~2, pp. 1131--1140, 2018.

\bibitem{RN272}
F.~Yang, X.~Song, G.~Dong, and K.-L. Tsui, ``A coulombic efficiency-based model
  for prognostics and health estimation of lithium-ion batteries,''
  \emph{Energy}, vol. 171, pp. 1173--1182, 2019.

\bibitem{RN398}
Z.~Long, Y.~Lu, X.~Ma, and B.~Dong, ``{PDE-Net}: Learning {PDEs} from data,''
  in \emph{Int. Conf. Mach. Learn.}\hskip 1em plus 0.5em minus 0.4em\relax
  PMLR, 2018, Conference Proceedings, pp. 3208--3216.

\bibitem{RN399}
Z.~Long, Y.~Lu, and B.~Dong, ``{PDE-Net} 2.0: Learning {PDEs} from data with a
  numeric-symbolic hybrid deep network,'' \emph{J. Comput. Phys.}, vol. 399, p.
  108925, 2019.

\bibitem{RN350}
M.~Raissi, ``Deep hidden physics models: Deep learning of nonlinear partial
  differential equations,'' \emph{J. Mach. Learn. Res.}, vol.~19, pp. 932--955,
  2018.

\bibitem{RN366}
S.~Cofre-Martel, E.~L. Droguett, and M.~Modarres, ``Remaining useful life
  estimation through deep learning partial differential equation models: A
  framework for degradation dynamics interpretation using latent variables,''
  \emph{Shock Vib.}, vol. 2021, p. 9937846, 2021.

\bibitem{RN479}
A.~Sherstinsky, ``Deriving the recurrent neural network definition and {RNN}
  unrolling using signal processing,'' in \emph{{CRACT Workshop at
  NeurIPS-2018}}, vol.~31, 2018, Conference Proceedings, p.~4.

\bibitem{RN359}
Y.~Yucesan and F.~Viana, ``A physics-informed neural network for wind turbine
  main bearing fatigue,'' \emph{Int. J. Prognostics Health Manag.}, vol.~11,
  no.~1, p.~17, 2020.

\bibitem{RN370}
Y.~A. Yucesan and F.~A.~C. Viana, ``Hybrid physics-informed neural networks for
  main bearing fatigue prognosis with visual grease inspection,'' \emph{Comput.
  Ind.}, vol. 125, p. 103386, 2021.

\bibitem{RN397}
Y.~Zhu, N.~Zabaras, P.~S. Koutsourelakis, and P.~Perdikaris,
  ``Physics-constrained deep learning for high-dimensional surrogate modeling
  and uncertainty quantification without labeled data,'' \emph{J. Comput.
  Phys.}, vol. 394, pp. 56--81, 2019.

\bibitem{RN289}
M.~Raissi, P.~Perdikaris, and G.~E. Karniadakis, ``Physics-informed neural
  networks: A deep learning framework for solving forward and inverse problems
  involving nonlinear partial differential equations,'' \emph{J. Comput.
  Phys.}, vol. 378, pp. 686--707, 2019.

\bibitem{RN402}
A.~Iserles, \emph{A first course in the numerical analysis of differential
  equations}.\hskip 1em plus 0.5em minus 0.4em\relax Cambridge Univ. Press,
  2009.

\bibitem{RN374}
J.~Shi, A.~Rivera, and D.~Wu, ``Battery health management using
  physics-informed machine learning: Online degradation modeling and remaining
  useful life prediction,'' \emph{Mech. Syst. Signal Process.}, vol. 179, p.
  109347, 2022.

\bibitem{RN474}
W.~D. Connor, Y.~Q. Wang, A.~A. Malikopoulos, S.~G. Advani, and A.~K. Prasad,
  ``Impact of connectivity on energy consumption and battery life for electric
  vehicles,'' \emph{{IEEE} Trans. Intell. Veh.}, vol.~6, no.~1, pp. 14--23,
  2021.

\bibitem{RN269}
J.~Kong, F.~Yang, X.~Zhang, E.~Pan, Z.~Peng, and D.~Wang, ``Voltage-temperature
  health feature extraction to improve prognostics and health management of
  lithium-ion batteries,'' \emph{Energy}, vol. 223, p. 120114, 2021.

\bibitem{RN57}
Z.~Liu, G.~Sun, S.~Bu, J.~Han, X.~Tang, and M.~Pecht, ``Particle learning
  framework for estimating the remaining useful life of lithium-ion
  batteries,'' \emph{{IEEE} Trans. Instrum. Meas.}, vol.~66, no.~2, pp.
  280--293, 2017.

\bibitem{RN388}
R.~Spotnitz, ``Simulation of capacity fade in lithium-ion batteries,'' \emph{J.
  Power Sour.}, vol. 113, no.~1, pp. 72--80, 2003.

\bibitem{RN252}
P.~Wen, S.~Zhao, S.~Chen, and Y.~Li, ``A generalized remaining useful life
  prediction method for complex systems based on composite health indicator,''
  \emph{Rel. Eng. Syst. Safety}, vol. 205, p. 107241, 2021.

\bibitem{RN354}
H.~N. Liu, I.~H. Naqvi, F.~J. Li, C.~L. Liu, N.~Shafiei, Y.~L. Li, and
  M.~Pecht, ``An analytical model for the {CC-CV} charge of {Li-ion} batteries
  with application to degradation analysis,'' \emph{J. Energy Storage},
  vol.~29, p. 101342, 2020.

\bibitem{RN385}
K.~A. Severson, P.~M. Attia, N.~Jin, N.~Perkins, B.~Jiang, Z.~Yang, M.~H. Chen,
  M.~Aykol, P.~K. Herring, D.~Fraggedakis, M.~Z. Bazant, S.~J. Harris, W.~C.
  Chueh, and R.~D. Braatz, ``Data-driven prediction of battery cycle life
  before capacity degradation,'' \emph{Nat. Energy}, vol.~4, no.~5, pp.
  383--391, 2019.

\bibitem{RN389}
S.~Zhao, Y.~Peng, F.~Yang, E.~Ugur, B.~Akin, and H.~Wang, ``Health state
  estimation and remaining useful life prediction of power devices subject to
  noisy and aperiodic condition monitoring,'' \emph{{IEEE} Trans. Instrum.
  Meas.}, vol.~70, pp. 1--16, 2021.

\bibitem{RN66}
K.~Liu, N.~Z. Gebraeel, and J.~Shi, ``A data-level fusion model for developing
  composite health indices for degradation modeling and prognostic analysis,''
  \emph{{IEEE} Trans. Autom. Sci. Eng.}, vol.~10, no.~3, pp. 652--664, 2013.

\bibitem{RN390}
P.~Wen, Y.~Li, S.~Chen, and S.~Zhao, ``Remaining useful life prediction of
  iiot-enabled complex industrial systems with hybrid fusion of multiple
  information sources,'' \emph{{IEEE} Internet Things J.}, vol.~8, no.~11, pp.
  9045--9058, 2021.

\bibitem{RN391}
K.~Hornik, M.~Stinchcombe, and H.~White, ``Multilayer feedforward networks are
  universal approximators,'' \emph{Neural Netw.}, vol.~2, no.~5, pp. 359--366,
  1989.

\bibitem{RN392}
A.~G. Baydin, B.~A. Pearlmutter, A.~A. Radul, and J.~M. Siskind, ``Automatic
  differentiation in machine learning: a survey,'' \emph{J. Mach. Learn. Res.},
  vol.~18, pp. 1--43, 2018.

\bibitem{RN463}
J.~Yu, L.~Lu, X.~Meng, and G.~E. Karniadakis, ``Gradient-enhanced
  physics-informed neural networks for forward and inverse pde problems,''
  \emph{Comput. Methods Appl. Mech. Eng.}, vol. 393, p. 114823, 2022.

\bibitem{RN377}
A.~Kendall, Y.~Gal, and R.~Cipolla, ``Multi-task learning using uncertainty to
  weigh losses for scene geometry and semantics,'' in \emph{Proc. {IEEE} Conf.
  Comput. Vision Pattern Recognit.}, 2018, Conference Proceedings, pp.
  7482--7491.

\bibitem{RN357}
S.~F. Wang, Y.~J. Teng, and P.~Perdikaris, ``Understanding and mitigating
  gradient flow pathologies in physics-informed neural networks,'' \emph{{SIAM}
  J. Scientific Comput.}, vol.~43, no.~5, pp. A3055--A3081, 2021.

\bibitem{RN376}
R.~Bischof and M.~Kraus, ``Multi-objective loss balancing for physics-informed
  deep learning,'' \emph{arXiv preprint: 2110.09813}, 2021.

\bibitem{RN291}
G.~Dong, W.~Han, and Y.~Wang, ``Dynamic bayesian network-based lithium-ion
  battery health prognosis for electric vehicles,'' \emph{{IEEE} Trans. Ind.
  Electron.}, vol.~68, no.~11, pp. 10\,949--10\,958, 2020.

\bibitem{RN342}
Y.~Ji, Z.~Chen, Y.~Shen, K.~Yang, Y.~Wang, and J.~Cui, ``An {RUL} prediction
  approach for lithium-ion battery based on {SADE-MESN},'' \emph{Appl. Soft
  Comput.}, vol. 104, p. 107195, 2021.

\bibitem{RN364}
S.~Pepe, J.~P. Liu, E.~Quattrocchi, and F.~Ciucci, ``Neural ordinary
  differential equations and recurrent neural networks for predicting the state
  of health of batteries,'' \emph{J. Energy Storage}, vol.~50, p. 104209, 2022.

\bibitem{RN379}
Y.-H. Lin and G.-H. Li, ``A bayesian deep learning framework for {RUL}
  prediction incorporating uncertainty quantification and calibration,''
  \emph{{IEEE} Trans. Ind. Informat.}, vol.~18, no.~10, pp. 7274--7284, 2022.

\bibitem{RN281}
M.~A. Patil, P.~Tagade, K.~S. Hariharan, S.~M. Kolake, T.~Song, T.~Yeo, and
  S.~Doo, ``A novel multistage support vector machine based approach for {Li}
  ion battery remaining useful life estimation,'' \emph{Appl. Energy}, vol.
  159, pp. 285--297, 2015.

\bibitem{RN478}
S.~Cuomo, V.~S. Di~Cola, F.~Giampaolo, G.~Rozza, M.~Raissi, and F.~Piccialli,
  ``Scientific machine learning through physics-informed neural networks: Where
  we are and what's next,'' \emph{J. Sci. Comput.}, vol.~92, no.~3, p.~88,
  2022.

\end{thebibliography}
\vfill\begin{IEEEbiography}[{\includegraphics[width=1in,height=1.25in,clip,keepaspectratio]{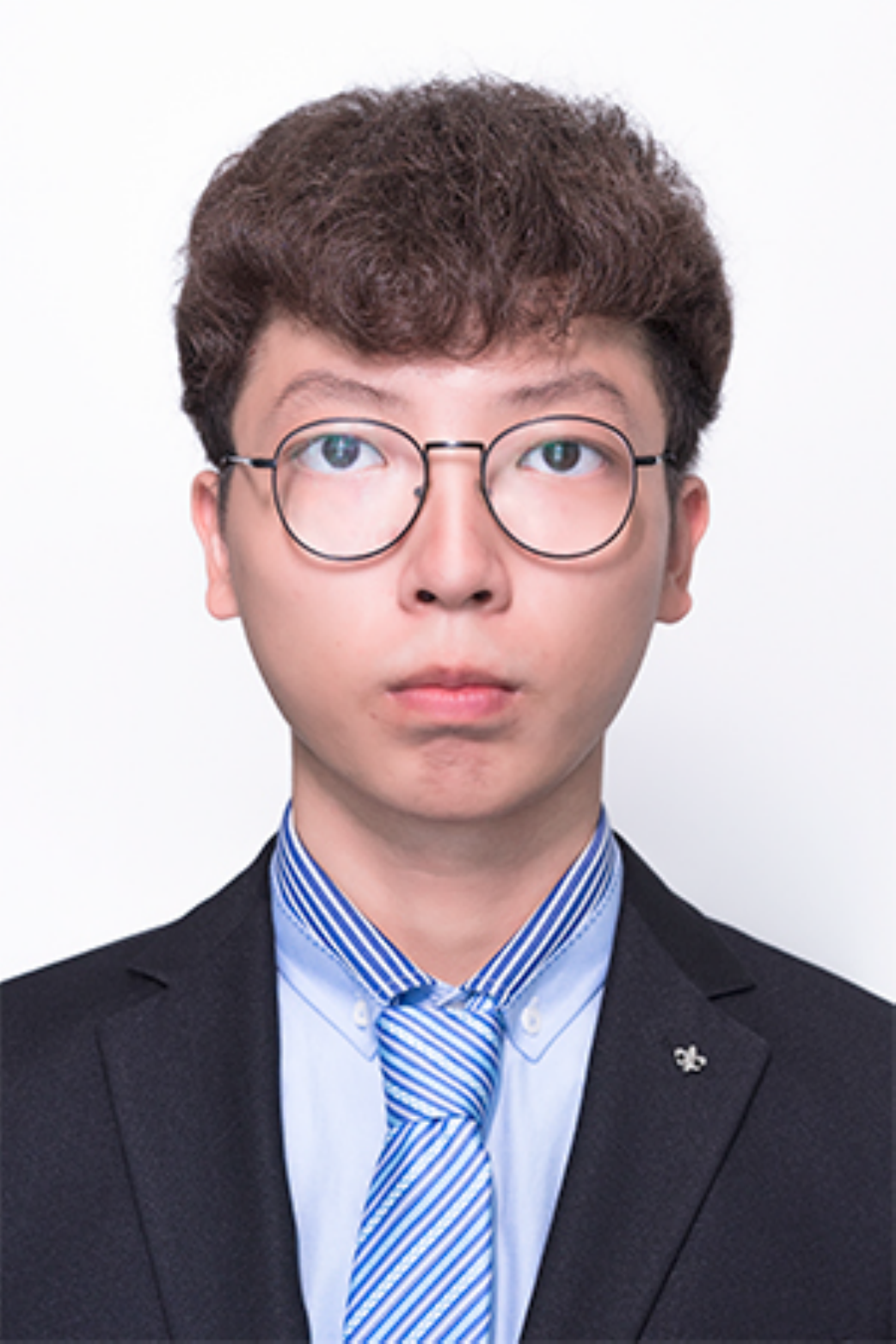}}]{Pengfei Wen} (S'20) received B.S. and Ph.D. degrees in Information and Communication Engineering from the School of Electronics and Information, Northwestern Polytechnical University (NWPU), Xi'an, China, in 2017 and 2023, respectively. From 2021 to 2022, he was a Visiting Ph.D. Student with the Department of Industrial Systems Engineering and Management, National University of Singapore, Singapore, with the scholarship from NWPU and Huawei Technologies Co., Ltd. His current research interests include Information Fusion, Reliability Engineering, and Physics-Informed Machine Learning.
\end{IEEEbiography}\vskip 0pt plus -1fil\begin{IEEEbiography}[{\includegraphics[width=1in,height=1.25in,clip,keepaspectratio]{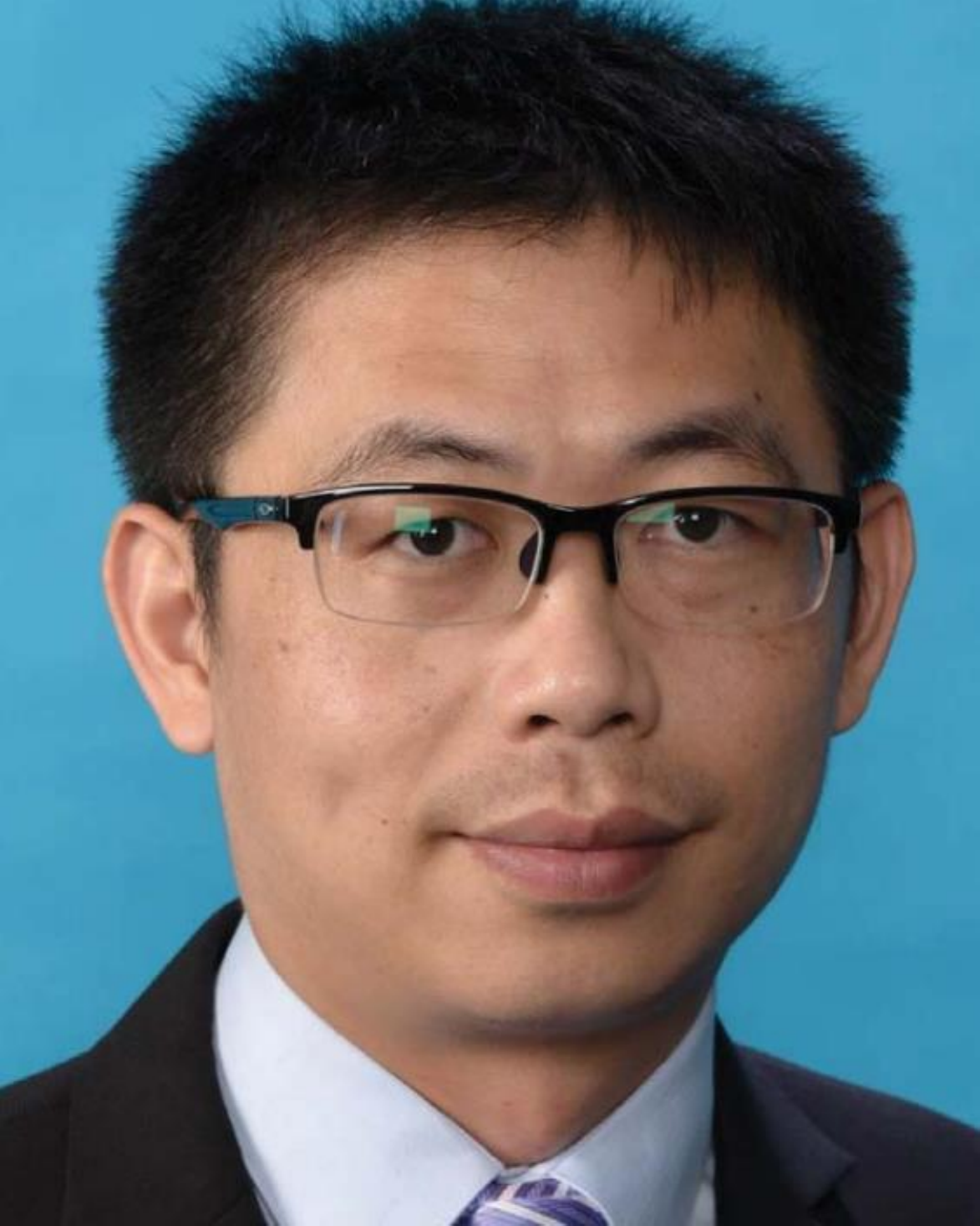}}]{Zhi-Sheng Ye} (Senior Member, IEEE) received the joint B.E. degree in material science \& engineering and in economics from Tsinghua University, Beijing, China, in 2008, and the Ph.D. degree in industrial and systems engineering from the National University of Singapore, Singapore, in 2012. He is currently an Associate Professor with the Department of Industrial Systems Engineering and Management, National University of Singapore. His research interests include reliability engineering, complex systems modeling, and industrial statistics.
\end{IEEEbiography}\vskip 0pt plus -1fil\begin{IEEEbiography}[{\includegraphics[width=1in,height=1.25in,clip,keepaspectratio]
{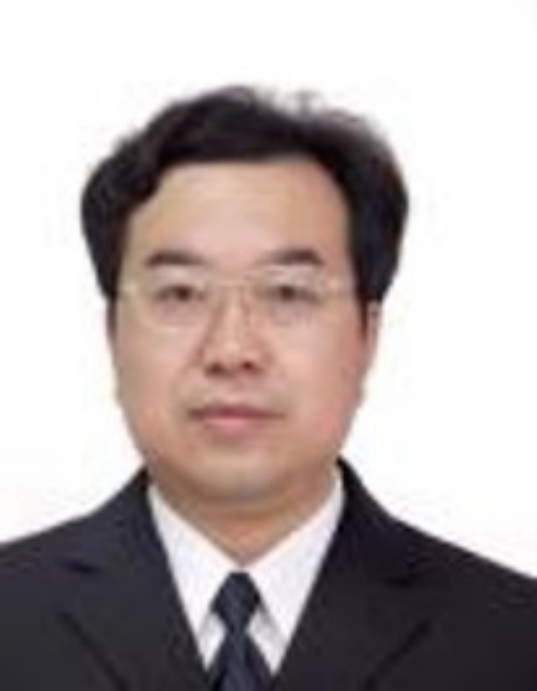}}]{Yong  Li}  received the B.S. degree in Avionics Engineering, M.S. and Ph.D degrees in   Circuits and Systems from Northwestern Polytechnical University, Xi'an,   China, in 1983, 1988 and 2005, respectively. He joined School of Electronic   Information, Northwestern Polytechnical University in 1993 and was promoted   to professor in 2002. His research interests include digital signal   processing and radar signal processing. \end{IEEEbiography}\vskip 0pt plus -1fil

\begin{IEEEbiography}[{\includegraphics[width=1in,height=1.25in,clip,keepaspectratio]{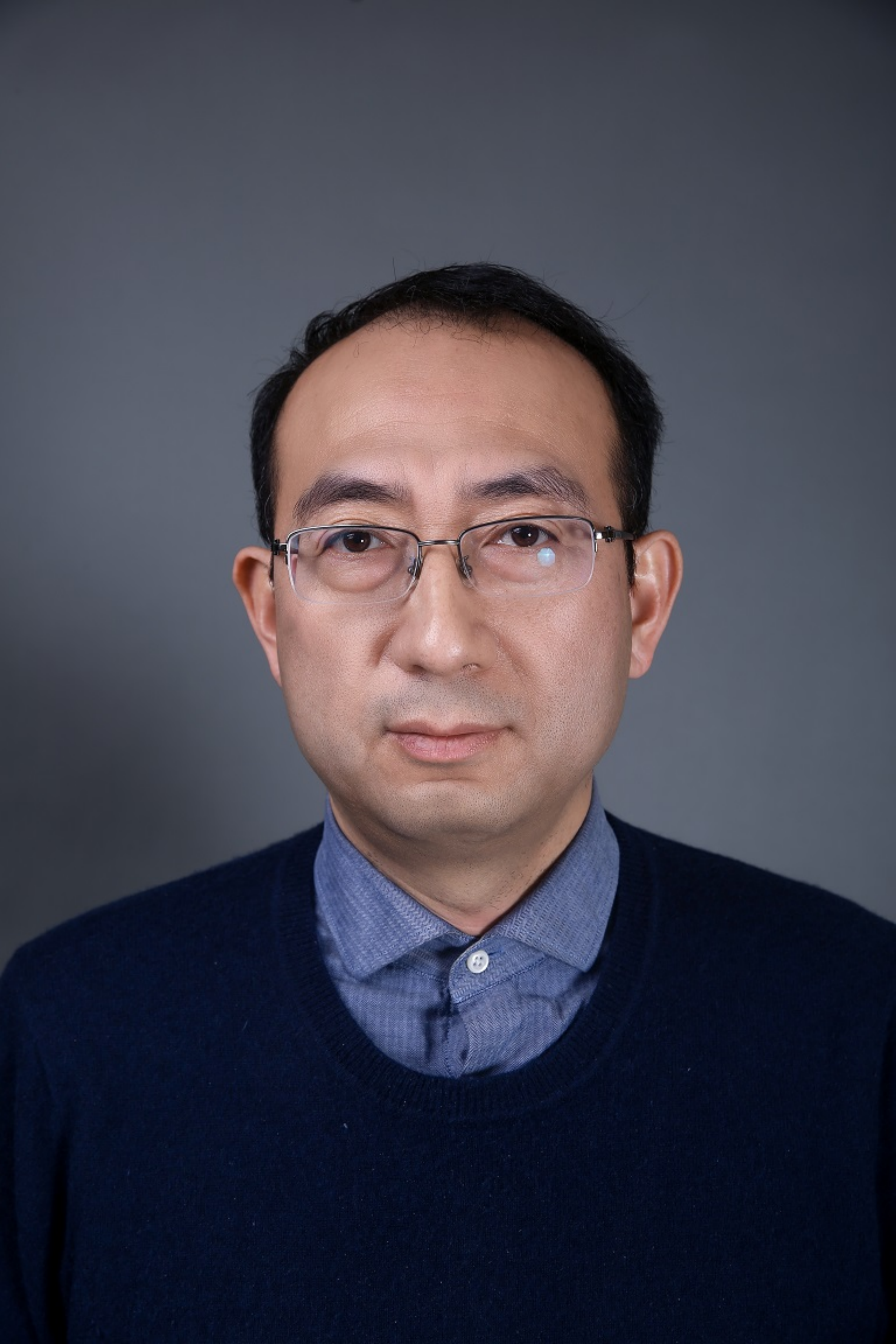}}]{Shaowei Chen} (M'15) is currently an associate professor in the School of Electronics and Information, Northwestern Polytechnical University, Xi'an, China, where he is also the dean of the Department of telcommunication engineering and the Director of Perception and IoT Information Processing Laboratory. He is the principal investigator of several projects supported by the Aeronautical Science Foundation of China, Beijing, China. His research expertise is in the area of the fault diagnosis, sensors, condition monitoring, and prognosis of electronic systems. Prof. Chen is selected to receive several Provincial and Ministerial Science and Technology Awards. He is a senior member of the Chinese Institute of Electronics.
\end{IEEEbiography}\vskip 0pt plus -1fil

\begin{IEEEbiography}[{\includegraphics[width=1in,height=1.25in,clip,keepaspectratio]{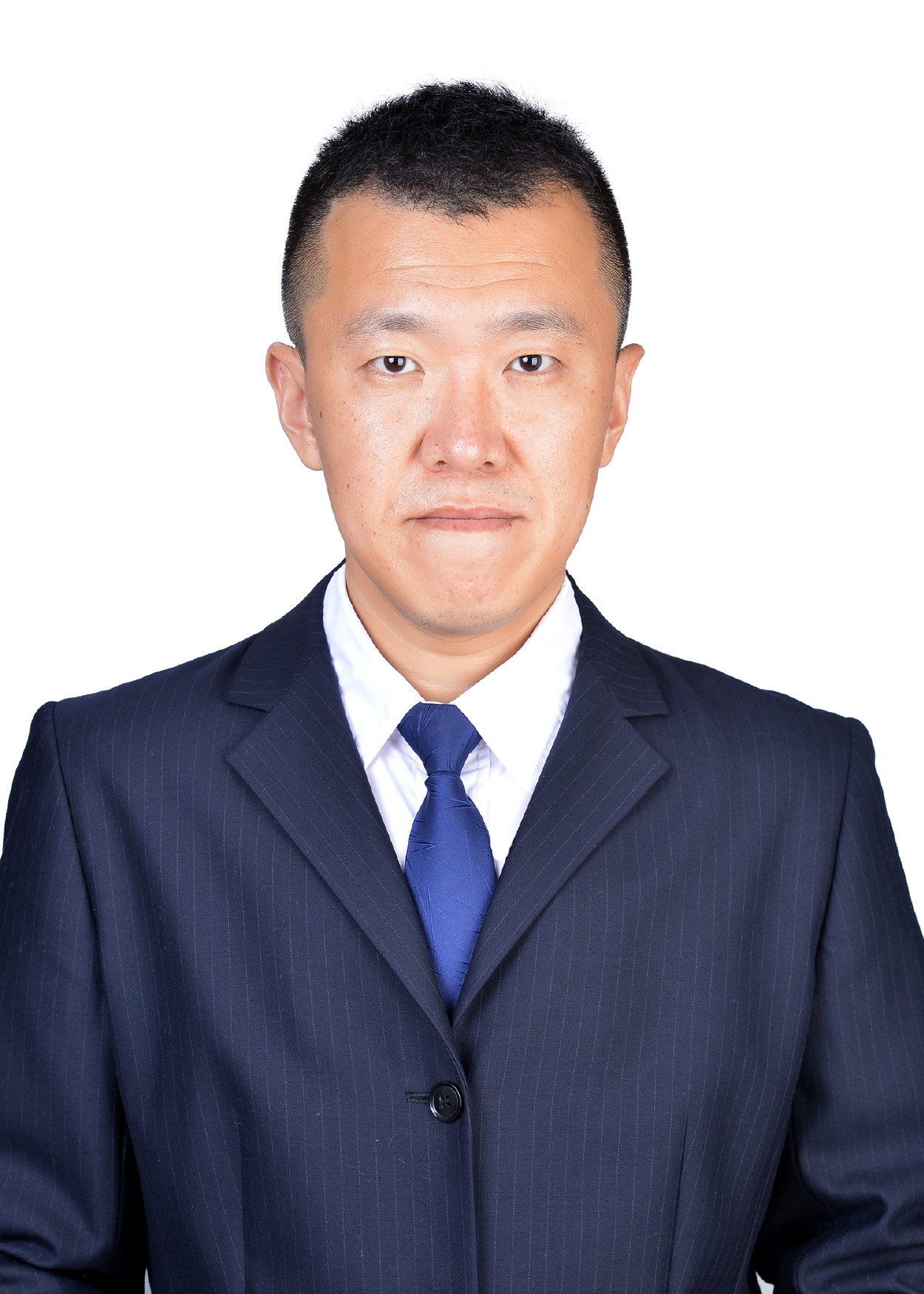}}]{Pu Xie}
obtained his PhD in the Department of Mechanical Engineering at Tsinghua University in 2016. After graduation, he joined CRRC Changchun Railway Vehicles Co., Ltd., where he conducted experimental research on structural fatigue and crack behavior using the MTS system. Currently, he is a visiting postdoctoral scholar in the Department of Aeronautics \& Astronautics at Stanford University. His research interests are focused on developing and implementing advanced testing and diagnostic methods that enhance the safety and reliability of mechanical systems, as well as developing multiphysics analytical methods for solving complex problems. 
\end{IEEEbiography}\vskip 0pt plus -1fil\begin{IEEEbiography}[{\includegraphics[width=1in,height=1.25in,clip,keepaspectratio]{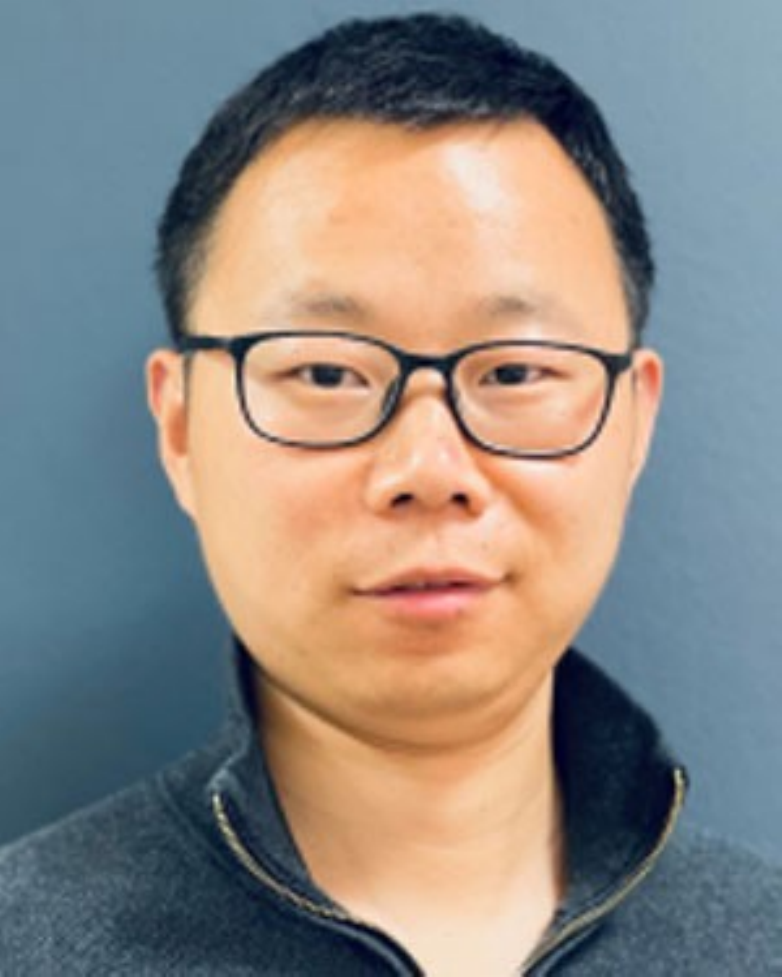}}]{Shuai Zhao}
(S'14-M'18) received the B.S., M.S., and Ph.D. degrees in information and telecommunication engineering from Northwestern Polytechnical University, Xi'an, China, in 2011, 2014, and 2018, respectively. He is currently an Assistant Professor with AAU Energy, Aalborg University, Denmark. From 2014 to 2016, he was a Visiting Ph.D. student with the Department of Mechanical and Industrial Engineering, University of Toronto, Canada. In August 2018, he was a Visiting Scholar with the Power Electronics and Drives Laboratory, Department of Electrical and Computer Science, University of Texas at Dallas, Richardson, TX, USA. From 2018 to 2022, he was a Postdoc researcher with AAU Energy, Aalborg University, Denmark. His research interests include physics-informed machine learning, system informatics, condition monitoring, diagnostics and prognostics, and tailored AI tools for power electronic systems. 
\end{IEEEbiography}\vskip 0pt plus -1fil 
\end{document}